\documentclass[preprint, 12pt, amsmath, amssymb, aps, superscriptaddress, showkeys]{revtex4-2}
\usepackage[utf8]{inputenc}
\usepackage[T1]{fontenc}        
\usepackage{slashed}
\usepackage{color}
\usepackage{graphicx}
\usepackage{bm}
\usepackage{diagbox}
\usepackage{booktabs}
\usepackage{multirow}
\usepackage{enumerate}
\graphicspath{{pics/}}
\allowdisplaybreaks[1]

\usepackage[
    colorlinks = true,
    linkcolor  = blue,
    citecolor  = blue,
    urlcolor   = blue
]{hyperref}

\begin{document}

\title{Probing azimuthal alignment in heavy-ion collisions: Clusterization effects}

\author{Aleksei Nikolskii}
\email{alexn@theor.jinr.ru}
\affiliation{Bogoliubov Laboratory of Theoretical Physics, Joint Institute for Nuclear Research, 141980 Dubna, Moscow Region, Russia}
\affiliation{Skobeltsyn Institute of Nuclear Physics, Lomonosov Moscow State University, 119991 Moscow, Russia}
\author{Igor Lokhtin}
\email{lokhtin@www-hep.sinp.msu.ru}
\affiliation{Skobeltsyn Institute of Nuclear Physics, Lomonosov Moscow State University, 119991 Moscow, Russia}
\author{Alexander Snigirev}
\email{snigirevam@my.msu.ru}
\affiliation{Bogoliubov Laboratory of Theoretical Physics, Joint Institute for Nuclear Research, 141980 Dubna, Moscow Region, Russia}
\affiliation{Skobeltsyn Institute of Nuclear Physics, Lomonosov Moscow State University, 119991 Moscow, Russia}

\begin{abstract}
The influence of kinematic constraints and event selection on the emergence of the alignment phenomenon observed in cosmic-ray experiments is studied within the HYDJET++ model, which is widely used for simulating heavy-ion collisions. It is demonstrated that the high degree of alignment, previously identified for realistic values of the transverse momentum disbalance of the most energetic particles, is also observed at the level of the most energetic clusters. In high-multiplicity heavy-ion collision events, the clustering procedure plays a crucial role in resolving individual particle groups on the detection plane, allowing a more accurate characterization of alignment patterns. These results highlight the combined effects of cluster formation and momentum conservation in shaping the observed azimuthal correlations.
\end{abstract}

\keywords{Properties of Hadrons; Specific QCD Phenomenology; Relativistic Heavy Ion Physics; Particle Correlations and Fluctuations; Quark-Gluon Plasma; Cosmic Rays}

\maketitle

\section{Introduction}
Angular correlations among secondary particles are a well-established feature of multiparticle production at both cosmic and accelerator energies. Among these, the alignment phenomenon represents a particularly striking case, first identified by the Pamir Collaboration in cosmic-ray emulsion experiments using large-area chambers placed at an altitude of about 5~km in the Pamir Mountains~\cite{pamir1,pamir2,pamir3,pamir4,Kopenkin:1994hu}. In these experiments, the most energetic hadrons and photons, or their clusters, tended to lie approximately along a straight line in the emulsion plane, suggesting a coplanar geometry of the events. Similar effects were later reported in the Capdevielle experiment aboard the Concorde aircraft~\cite{Capdevielle:1988pe}. Despite considerable theoretical effort~\cite{Capdevielle:1988pe,Kopenkin:1994hu,Halzen:1989rg,Kopenkin:1994yu,Mukhamedshin:2005nr,Lokhtin:2005bb,Lokhtin:2006xv,DeRoeck:2010voa,Dremin:2010yd}, a universally accepted explanation of this phenomenon has not yet been achieved.

Evidence for alignment in collider experiments remains lacking, even though the collision energies at the Large Hadron Collider (LHC) exceed the effective threshold $\sqrt{s_{\rm eff}} \gtrsim 4$~TeV observed in cosmic-ray interactions. At the same time, other forms of long-range azimuthal correlations -- such as the ridge effect first observed by the STAR collaboration~\cite{STAR:2005ryu,STAR:2009ngv} later at RHIC~\cite{PHOBOS:2008yxa} and in high-multiplicity proton-proton collisions at the LHC~\cite{CMS:2010ifv} -- have been extensively studied. Several works have attempted to link these effects to alignment~\cite{Dremin:2010yd,Lokhtin:2013pva,Mukhamedshin:2022eek}, but differences in kinematic conditions and reference frames prevent any direct correspondence. Moreover, the ridge structure can be successfully described within conventional hydrodynamic and flow-based frameworks~\cite{Eyyubova:2014dha}, supporting the interpretation of alignment as a statistical fluctuation and phenomenological models~\cite{Mukhamedshin:2005nr,Mukhamedshin:2019dus}.

In our previous analyses~\cite{Lokhtin:2023tze,Lokhtin:2024trs}, we proposed a geometric interpretation of alignment, demonstrating that pronounced collinearity can naturally arise due to the selection procedure of the most energetic particles, the energy-deposition threshold, and transverse momentum conservation. This concept was later implemented in the realistic heavy-ion event generator HYDJET++~\cite{Lokhtin:2008xi}, whose statistical model allows for event-by-event conservation of global quantities such as total transverse momentum and net charge \cite{Lokhtin:2024sbm}.

The present study extends this approach by examining the impact of particle clustering in a high-multiplicity environment on the degree of alignment. The remainder of the paper is organized as follows. Section~\ref{sec:notation} introduces the essential notation for alignment analysis. Section~\ref{sec:alignment_v2} provides an overview of the alignment phenomenon and elliptic flow. Section~\ref{sec:simulation} describes the simulation of alignment in heavy-ion collisions. Section~\ref{sec:pt_conservation} discusses the influence of event-by-event transverse momentum conservation. Section~\ref{sec:comparison} compares simulations with and without clustering under transverse momentum conservation. Section~\ref{sec:conclusion} provides concluding remarks. Additional details on modeling statistics are provided in the Appendix.

\section{Essential notation for Alignment Analysis} 
\label{sec:notation}

In the Pamir experiment, families were selected and analyzed in which the total energy of $\gamma$ quanta exceeded a certain threshold and in which at least one hadron was present. The alignment effect becomes pronounced at $\sum E_{\gamma} > 0.5$~PeV, corresponding to interaction energies of $\sqrt{s} \gtrsim 4$~TeV. These families are produced mainly by protons with energies $\gtrsim 10^4$~TeV interacting at an altitude $h$ ranging from several hundred meters to a few kilometers in the atmosphere above the chamber~\cite{pamir1,Kopenkin:1994hu}. The collision products are observed within a radial distance $r_{\rm max}$ up to about 15~mm in the emulsion, with a minimum spot separation $r_{\rm min} \sim 1$~mm. 

The alignment parameter $\lambda_N$ quantifies the azimuthal correlation among $N$ selected particles or clusters and characterizes their deviation from a straight line, providing a more sensitive measure of asymmetry than other parameters, such as eccentricity or thrust. It is a dimensionless quantity taking values in the interval $\left[-1/(N-1),1\right]$ and is defined as~\cite{pamir4},
\begin{equation}
\label{eq:lambda}
\lambda_{N}~=~ \frac{ \sum^{N}_{i \neq j \neq k}\cos(2 \varphi_{ijk})}
{N(N-1)(N-2)},
\end{equation}
where $\varphi_{ijk}$ is the angle between two straight lines connecting the $i$th spot with the $j$th and $k$th spots. The combinatorial factor $N(N-1)(N-2)$ accounts for all triplet combinations and ensures proper normalization. For example, if $N=3$, an equilateral triangle yields $\lambda_3=-0.5$. Perfect alignment of all points along a straight line corresponds to $\lambda_N=1$, independent of $N$, while deviations from a straight line reduce $\lambda_N$, reflecting a lower degree of collinearity.

The degree of alignment
\begin{equation}
\label{deg_align}
P_N=\frac{l}{L} 
\end{equation}
is defined as the ratio of events $l$ for which $\lambda_N > 0.8$ to the total number of events $L$ and only events with at least $N$ energy centers are counted.~\cite{Kopenkin:1994hu}.

For convenience  we parametrize the four-momentum of each particle $i$ under consideration with its transverse momentum $p_{Ti}$ (with respect to the collision axis $z$), mass $m_i$, rapidity $\eta_i$ and  azimuthal angle $\phi_i$ in the center-of-mass system:
\begin{widetext}
\begin{equation}
\label{momentum}
[\sqrt{p^2_{Ti}+m^2_i}~\cosh \eta_i,~~~ p_{Ti}\cos \phi_i,~~~ p_{Ti}\sin \phi_i,~~~ \sqrt{p^2_{Ti}+m^2_i}~\sinh \eta_i].
\end{equation}
\end{widetext}

The transformation from the center-of-mass system to the laboratory frame  amounts to the rapidity shift: 
\[
\zeta_i = \eta_0 + \eta_i,
\]
where $\eta_0$, $\zeta_i$ are the rapidities of the center-of-mass system and the particle $i$, respectively, in the laboratory reference frame.

\begin{figure} 
\begin{center}
\includegraphics[scale=0.6]{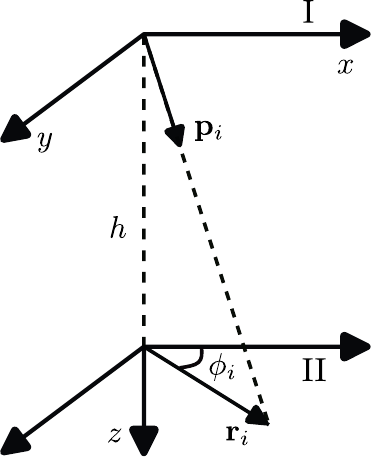}
\caption{Kinematics and scheme of the analyzed events in the context of Pamir experiment.
Region~I represents the Earth’s atmosphere, where incoming cosmic-ray protons generate cascades of hadrons and photons whose alignment properties are studied.
Region~II corresponds to the emulsion film plane.
In this region, $p_i$ stands for the momentum of a registered particle (or cluster), $r_i$ denotes its position on the film, $\phi_i$ is the azimuthal angle, and $h$ indicates the height above the emulsion chamber at which the particle was produced.}
\label{fig:kinematics}
\end{center}
\end{figure}	

If one neglects the further interactions of the particles propagating through the atmosphere, this provides an upper estimate of the alignment effect, then 
the needed azimuthal angles (Fig. \ref{fig:kinematics}) are calculated over the positions ${\bf r}_i$ of the particles in the $(x-y)$ plane in the film,
\begin{equation}
\label{position}
{\bf r}_i~=~ \frac{{\bf v}_{ri}}{v_{zi}}~h,
\end{equation}
where  $v_{zi}$ and  ${\bf v}_{ri}$ are the longitudinal and radial components of particle velocity respectively. Accounting for the transformation from the center-of-mass frame to the laboratory frame via a rapidity shift, the particle coordinates are given by:
\begin{equation}
\label{eq:positionl}
{\bf r}_i~=~\frac{{\bf p}_{Ti}}{\sqrt{p^2_{Ti}+m^2_i}~\sinh (\eta_0+\eta_i)}~h.
\end{equation}

Since the size of the observation region is about several centimeters, these radial distances $r_i=\left| {\bf r}_i \right|$ must obey the following restrictions:
\begin{equation}
\label{eq:rmin}
 r_{\rm min}~<~r_i,
\end{equation}
\begin{equation}
\label{eq:rmax}
 r_{i}~<~r_{\rm max}.
\end{equation}
The condition \eqref{eq:rmin} simply means that spots are not mixed with the
central one formed by the particles that fly close to the collision axis $z$, predominantly region of incident-hadron fragmentation, and the condition \eqref{eq:rmax} indicates that the particle coordinate does not exceed the observation region. 

In this context, it is also interesting and relevant to consider the influence of the applied restrictions $r_{\rm min} < r_i < r_{\rm max}$ (the laboratory acceptance criterion) on the particle spectrum selected for the alignment calculation. For particles with sufficiently large transverse momenta $p_{Ti}$ compared to their masses $m_i$, the conditions \eqref{eq:rmin}, \eqref{eq:rmax} reduce mainly to restrictions on the available particle rapidities in the center-of-mass system,
\[
r_{\rm min} < r_i \;\Longrightarrow\; \eta_i < \eta_{\rm max} = \ln (r_0 / r_{\rm min}) \simeq 5.98 ,
\]
\[
r_i < r_{\rm max} \;\Longrightarrow\; \eta_i > \eta_{\rm min} = \ln (r_0 / r_{\rm max}) \simeq 3.28 ,
\]
since in this case $r_i \simeq r_0 / e^{\eta_i}$ for $\eta_0 + \eta_i \gtrsim 1$, where $r_0 = 2h / e^{\eta_0}$. In this estimation, we have set $h=1$ km, $r_{\rm min}=1$ mm, $r_{\rm max}=15$ mm, and $\eta_0=8.52$ (at $\sqrt{s}=5.02$ TeV), for the sake of clarity.

Among clusters that satisfy the restrictions \eqref{eq:rmin}, \eqref{eq:rmax} one selects the two to seven clusters/particles that are most energetic. After that one
calculates the alignment $\lambda_N$ using the definition above and taking into account the central cluster, i.e., $N-1=2\ldots7$. 

\section{On Alignment and Azimuthal Flow}
\label{sec:alignment_v2}

\begin{figure*}[t]
\begin{center}
\includegraphics[scale=0.8]{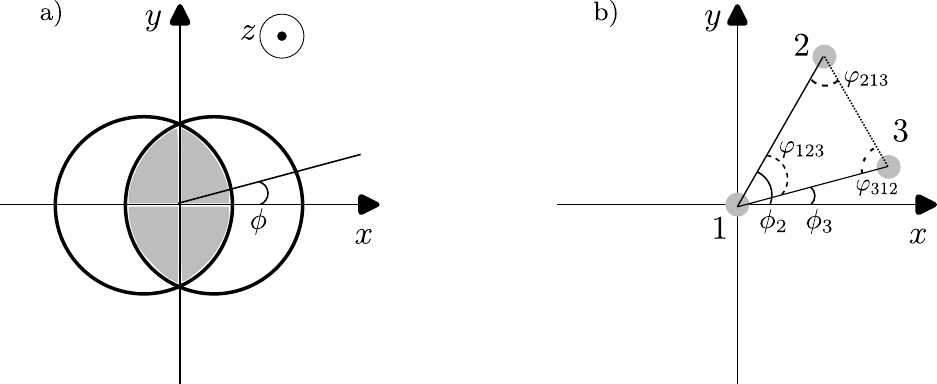}
\caption{
(a) Schematic illustration of a noncentral collision of two nuclei in the $x-y-z$ plane; $\phi$ denotes the azimuthal angle of the outgoing secondary particles.  
(b) Illustration of three clusters in the azimuthal plane used to study the alignment effect.  
The angles $\phi_1 = 0$, $\phi_2$, and $\phi_3$ correspond to the azimuthal positions of clusters 1, 2, and 3, respectively.  
The quantities $\varphi_{123}$, $\varphi_{213}$, and $\varphi_{312}$ represent the angles between the vectors connecting the clusters and are used in the alignment calculation (see Eq.~\eqref{eq:lambda}).  
}
\label{fig:v2-al}
\end{center}
\end{figure*}	

After introducing the essential parameters and observables relevant for the characterization of alignment, it is instructive to discuss the strengths and limitations of this method in comparison with the well-established description of azimuthal flow in heavy-ion collisions based on the Fourier harmonic decomposition. Historically, the alignment-based characterization of azimuthal correlations--originally developed in cosmic-ray studies--preceded the now-standard Fourier approach. The method based on the Fourier decomposition of the continuous distribution of outgoing particles over the azimuthal angle was introduced in Ref.~\cite{Voloshin:1994mz} and relies on the expression 
\begin{widetext}
\begin{equation}
\label{eq:Fourier}
E\frac{d^3N}{d^3p}=\frac{d^3N}{p_Tdp_Td\eta d\phi}=\frac{d^2N}{p_Tdp_Td\eta 2\pi} \Big[ 1+\sum^{\infty}_{n=1}2v_n\cos n(\phi-\Psi_R) \Big], 
\end{equation}
\end{widetext}
where $E$ denotes the particle energy, $p_T$ the transverse momentum, $\eta$ the rapidity, $\phi$ the azimuthal angle of outgoing particles, and $\Psi_R$ the reaction-plane angle. Figure~\ref{fig:v2-al}(a) schematically illustrates a noncentral collision of two nuclei and the direction of a secondary particle emitted at an azimuthal angle $\phi$. Since the reaction-plane angle is difficult to determine experimentally, it is usually replaced by the so-called event-plane angle; for simplicity, in Fig.~\ref{fig:v2-al}(a) we set $\Psi_R=0$. The first two coefficients, $v_1$ and $v_2$, correspond to directed and elliptic flow, respectively, while the leading term describes the azimuthally symmetric component of the particle yield. From the orthogonality of the Fourier basis, one obtains
\begin{widetext}
\begin{equation}
\label{vn}
v_n(p_T,\eta)=\langle \cos \big[n(\phi-\Psi_R)\big] \rangle = \frac{ \int_0^{2\pi}~d\phi \cos \big[n(\phi-\Psi_R)\big]\frac{d^3N}{p_Tdp_Td\eta d\phi} }{ \int_0^{2\pi}~d\phi \frac{d^3N}{p_Tdp_Td\eta d\phi} }, 
\end{equation}
\end{widetext}
where the averaging is first performed over all particles in a given event and subsequently over events within a centrality class. The second coefficient, $v_2(p_T,\eta)$,
has received the most attention and is the best studied, as it is strongly driven by the ellipticity of the nuclear overlap region and reflects the preference for particle emission along the short axis of the almond-shaped geometry (Fig.~\ref{fig:v2-al}(a)). However, there is now substantial evidence for sizeable contributions from higher-order harmonics as well~\cite{STAR:2022ncy,Devi:2023wih,ALICE:2024fus}.

Let us now turn to Fig.~\ref{fig:v2-al}(b), which schematically illustrates three clusters ($N=3$, with cluster~1 fixed at the origin for simplicity) in the azimuthal plane used to study alignment.
A fundamental difference from the Fourier-harmonic method \eqref{eq:Fourier} is that the alignment observable represents a nonflow type of particle correlation. In other words, it does not require particles to share similar $p_T$, rapidity, or azimuthal angle $\phi$. Instead, it focuses on the relative orientation of neighboring or nearby particle clusters, i.e., the angles between the clusters 1--3: $\varphi_{123}$, $\varphi_{213}$, $\varphi_{312}$ in Fig.~\ref{fig:v2-al}(b). In this sense, the alignment procedure may be viewed as a more general characterization of azimuthal structure, as it directly probes the mutual spatial distribution of particle flows in the azimuthal plane.

A natural question arises: is it possible to compare the approaches used to study azimuthal anisotropy through the Fourier coefficient $v_n$ and through particle alignment? Are there any correlations between these two observables, and can they be meaningfully combined in heavy-ion collisions? Strictly speaking, a direct comparison is essentially impossible, since, as noted above, $v_2$ (and higher harmonics) quantify collective flow, whereas alignment represents a nonflow particle correlation. In this context, it is more reasonable to introduce a nonflow quantity 
$c_2 \{2\}$, often referred to as a two-particle correlator~\cite{Voloshin:1994mz} or cumulant,
\begin{equation}
\label{eq:v2cor}
c_2 \{2\}= \langle \cos \left[ 2(\phi_i - \phi_j) \right] \rangle = \frac{1}{N_{\rm pairs}} \sum_{i < j}^N \cos \left[ 2(\phi_i - \phi_j) \right],
\end{equation}
where $N_{\rm pairs}=N(N-1)/2$ is the number of particle pairs constructed from $N$ particles, and $\phi_i$, $\phi_j$ denote the azimuthal angles of the $i$th and $j$th particles. In this form, 
$c_2\{2\}$ is the closest nonflow analog of ``classical'' $v_2$ and is therefore the most directly comparable to alignment. It is worth noting that the correlator method has been widely used in the analysis of relativistic heavy-ion collisions, as documented, for example, in Refs.~\cite{Poskanzer:1998yz,PHOBOS:2007vdf,Jia:2012ma,Bhalerao:2013ina,Busza:2018rrf,Bravina:2020sbz} and $c_2\{2\}=v_2^2$ if particles correlate with the reaction plane only. The methods discussed--the alignment $\lambda$, the flow $v_n$ decomposition technique \eqref{eq:Fourier}, and the correlator-based approach $c_n\{2\}$--can provide complementary information on azimuthal anisotropies in collisions of relativistic nuclei. 

The alignment method is particularly sensitive to nonflow effects and captures the relative geometric arrangement of nearby particle clusters, offering a direct probe of local structures in the azimuthal plane. Its main limitation lies in reduced statistical stability--this effect has not yet been observed in collider experiments, as well as the lack of a direct connection to the global collective flow. In contrast, the Fourier-harmonic decomposition for the elliptic flow coefficient $v_2$ and the correlator $c_2\{2\}$ are well-established, statistically more robust observables directly related to collective dynamics, although they may mix flow and nonflow contributions unless additional correction procedures are applied. It is important to note that both the flow $v_2$ and nonflow $c_2\{2\}$ (in the absence of particle correlations) quantities vanish for an isotropic azimuthal distribution. The absence of direct evidence for alignment in collider experiments, coupled with the established observation of elliptic flow through $v_2$ and $c_2\{2\}$ in the same experiments, strongly motivates the investigation of possible azimuthal alignment in heavy-ion collisions.

\section{Simulation of alignment in heavy-ion collisions}
\label{sec:simulation}
To obtain a clearer picture of how azimuthal alignment may appear under controlled conditions, we extended our earlier analysis \cite{Lokhtin:2023tze,Lokhtin:2024sbm} by defining three centrality selections: $0\%$--$5\%$, $40\%$--$75\%$, and the inclusive $0\%$--$75\%$ sample. The most central class ($0\%$--$5\%$) corresponds to near-head-on collisions and yields the largest particle multiplicities. The peripheral class ($40\%$--$75\%$) represents collisions in which anisotropic flow effects are expected to be most pronounced. The inclusive $0\%$--$75\%$ sample combines events over a broad range of collision geometries and centralities and therefore provides a more suitable basis for comparison with cosmic-ray data, in particular with the Pamir experiment. Although such a comparison is necessarily indirect and partly driven by the need to relate collider simulations to cosmic-ray observations, it nevertheless offers a useful reference framework for interpreting possible alignment effects.

Our goal is to examine the possible emergence of alignment structures in simulated heavy-ion reactions while consistently accounting for the system's space-time evolution and the balance between soft and hard particle production mechanisms. The comparison among different centrality classes also provides a way to study how alignment features may depend on the impact parameter of the initial interaction.

For the simulations, we employ the HYDJET++ event generator \cite{Lokhtin:2008xi}, a model validated against multiple nucleus-nucleus collision observables at RHIC and LHC energies. Recent developments are summarized in~\cite{Chernyshov:2022oik,Ambaryan:2024jhe,Ambaryan:2025lhy}. This framework allows one to model particle production in a realistic environment that includes thermal emission as well as jet-related contributions. The simulation accounts for soft particles with relatively low transverse momenta as well as the hard, jet-producing part of the generator, thus providing a comprehensive description of the collision dynamics.

In our analysis, we focus on Pb+Pb collisions at a center-of-mass energy of $\sqrt{s} = 5.02$  TeV per nucleon pair, using the HYDJET++ model with parameters taken from previous calibrations without additional retuning. This system provides a well-established testing ground for studying collective effects in strongly interacting matter. Although Pb+Pb collisions involve a much larger system than those produced in interactions of cosmic nuclei in the atmosphere, such as Fe+O, they can be explored in detail at modern colliders and are reliably described by existing Monte Carlo generators. The alignment phenomenon may, however, originate from nucleus--nucleus interactions at high energies. The collision energy at the LHC far exceeds the energy scale relevant for particle interactions in the Pamir experiment where this effect was observed. Furthermore, the formation of a quark-gluon plasma (QGP) requires surpassing a certain energy density, which is typically achieved in heavy-ion collisions. Nevertheless, a large atomic number is not a strict requirement for QGP formation; such a dense medium can also be produced in smaller systems, even in high-multiplicity proton-proton collisions~\cite{Pasechnik:2016wkt,PHENIX:2018lia,Sahoo:2023tmn}. Thus, while the consideration of Pb+Pb collisions is not essential for the qualitative features of our findings, this system was chosen for the present study because it has been extensively investigated experimentally and accurately reproduced by the HYDJET++ model without analysis-specific parameter retuning.

\subsection{Brief overview of the HYDJET++ model}
HYDJET++ is a Monte Carlo event generator designed for the simulation of relativistic heavy-ion collisions. It combines two main components:
\begin{itemize}
	\item The soft part represents a thermalized hadronic medium emerging at chemical and thermal freeze out, modeled via relativistic hydrodynamics with parametrized freeze-out conditions. Particle yields are sampled from a Poisson distribution, with the mean proportional to the number of participant nucleons. Collective flow and spatial anisotropies are implemented using the modified FAST MC generator \cite{Amelin:2006qe,Amelin:2007ic}, while the effective thermal volume accounts for flow profiles and freeze-out geometry, making particle ratios relatively insensitive to detailed freeze-out specifications.
	\item The hard part describes high-$p_T$ multiparton production and energy loss in the quark-gluon plasma, following the PYQUEN approach \cite{Lokhtin:2005px}, which accounts for both collisional and radiative mechanisms. The number of generated jets is drawn from a binomial distribution, with the mean determined by the number of binary nucleon-nucleon interactions and the hard-process cross section above a minimum transverse momentum $p_T^{\rm min}$. Partons below this threshold are considered thermalized and included in the soft sector.
\end{itemize}

It is important to note that this model does not feature an explicit quark-gluon plasma. Its effects are effectively incorporated through parton hadronization with a characteristic energy scale, which should be understood as approximate rather than strict. This scale reflects the conditions under which the azimuthal alignment phenomenon may emerge, signaling the onset of specific dynamical regimes in high-energy heavy-ion collisions. 

\subsection{Modeling approach using secondary-particle clustering} 
In the initial approach, each particle is treated as an individual cluster with variable size and mutual separation, limited only by the conditions \eqref{eq:rmin} and \eqref{eq:rmax} to prevent overlap with the central region of the detection plane. This simplified treatment does not account for the possibility that closely spaced particles may merge into a single, compound cluster; results from this procedure were discussed in our previous work~\cite{Lokhtin:2024sbm}.

In the present study, we explicitly incorporate clustering of secondary particles and the simulation procedure can be summarized as follows:
\begin{enumerate}[i.]
	\item First, an event corresponding to a collision of two nuclei of a given type is generated; 
	\item For each secondary particle, its position $r_i$ on the emulsion plane is calculated using Eq.~\eqref{position}, and it is verified whether the resulting position satisfies the acceptance criteria in Eqs.~\eqref{eq:rmin} and~\eqref{eq:rmax};
	\item After determining the position of a particle $i$, the distance to another particle $j$ is computed using the expression,
		\begin{equation}
		\label{eq:distance}
		d_{ij}~=~\sqrt{r^2_i~+~r^2_j~-~2r_i r_j \cos(\phi_i~-~\phi_j)} ;
		\end{equation}		
	\item If the distance $d_{ij}$ satisfies the clustering condition 
		\begin{equation}
		\label{clust_cond}
		d_{ij}<r_\text{res},
	\end{equation}
	with a given resolution parameter $r_\text{res}$, the particles are merged into a new cluster, whose coordinates are defined as 
		\begin{equation}
		\label{eq:cluster_pos}
		{\bf r}_{ij}=({\bf r}_i E_i+ {\bf r}_j E_j)/(E_i+E_j).
	\end{equation}
 	Otherwise, they are retained as separate clusters;
	\item Among the accepted clusters the conditions Eqs.~\eqref{eq:rmin}, \eqref{eq:rmax} are checked again. Then, three to five ($N=3$–$5$) highest-energy clusters are selected, and their alignment $\lambda_N$ and degree of alignment $P_N$ are computed according to Eqs.~\eqref{eq:lambda} and~\eqref{deg_align}. The most energetic cluster is always fixed at the origin \( O(0,0) \).
\end{enumerate}

This algorithm involves pairwise comparisons of all secondaries within an event and is conceptually analogous to standard jet-finding procedures. The main idea behind pairwise clustering algorithms is the concept of a ``nearest neighbor'' and the choice of a measure used to define this ``nearestness.'' In our case, the measure is the ordinary distance in the ``emulsion plane'' (using the terminology of the Pamir experiment). In collider experiments, typical measures for jet reconstruction include the square of the invariant mass, the transverse momentum of particles relative to the direction defined by the sum of two particles' momenta, or other more complex variables in phase space.

In the context of the alignment phenomenon in heavy-ion collisions, the particle positions $r_i$ given by Eq.~\eqref{eq:positionl}, together with the distance and clustering criteria defined in Eqs.~\eqref{eq:distance}--\eqref{eq:cluster_pos}, are used to characterize the spatial relations of particles in the azimuthal plane. For clarity, we note the following analogy: in jet physics, a widely used expression for jet selection is
\[
\Delta R_{\rm jet} = \sqrt{(\Delta \eta)^2 + (\Delta \phi)^2},
\]
where $\eta$ is the rapidity and $\phi$ is the azimuthal angle. In our case, a similar expression can be written as
\begin{equation}
\label{deldfull}
\Delta d = r \, \sqrt{(\Delta \eta)^2 + (\Delta \phi)^2},
\end{equation}
which is analogous to Eq.~\eqref{eq:distance} when expressed in terms of $\eta$ and $\phi$.

The Eqs.~\eqref{eq:positionl},~\eqref{deldfull} allows us to establish a relationship between the variation of the distance between clusterlike particles, $\Delta d$, in the laboratory frame and the variations of the center-of-mass variables $\eta$ and $\phi$, namely,
\[
\Delta d = r\,\Delta \phi
\quad \text{and} \quad
\Delta d = - r\,\Delta \eta ,
\]
where $r = |{\bf r}_i|$. The latter relation can be derived if $\sinh (\eta_0+\eta_i)$ is approximated by $\exp (\eta_0+\eta_i)/2$ for sufficiently large $(\eta_0+\eta_i)$. In this case, the clustering condition \eqref{clust_cond} can be rewritten as
\[
\sqrt{ (\Delta \eta)^2 + ( \Delta \phi)^2 } < r_{\rm res}/r ,
\]
where the connection between the ordinary (laboratory-frame) and center-of-mass coordinates becomes more transparent.

It is therefore possible to estimate the effective resolution in the center-of-mass variables for alignment clusters. For example, taking $r_{\rm res} = 1$~mm and $r = r_{\rm max} = 15$~mm, the resolution can be estimated as
\[
\eta_{\rm res} = \phi_{\rm res} = \frac{1}{15}.
\]
If instead $r = r_{\rm min} = 1$~mm, one obtains
\[
\eta_{\rm res} = \phi_{\rm res} = 1.
\]
This estimate is independent of the height $h$ and indicates that the separation between distinguishable clusters in $\eta$ and $\phi$ lies in the range
\[
\Delta \eta,\, \Delta \phi \in [1/15,\,1],
\]
depending on the radial position $r_i$ of the cluster. If the rapidity or azimuthal angle difference between clusters exceeds this resolution scale, they are treated as distinct; otherwise, they are merged into a single cluster.

Such measures are commonly used in modern event generators to identify jets in electron-positron annihilation and hadron collisions. The efficiency (i.e., the number of iterative steps) and the stability of clustering algorithms depend not only on the choice of the measure itself, but also on its numerical value and on the selection of particle pairs from which the clustering procedure is initiated. The clustering algorithm presented in this work demonstrates stability with respect to these criteria. Alignment observables are then calculated using only those clusters (or clusterlike particles) that do not meet the clustering condition in Eq.~\eqref{clust_cond}.

\subsection{Simulation outcomes with clustering of secondary particles}
The simulation results obtained according to the procedure described above, as well as the data from the Pamir experiment, are presented in Table \ref{tab:results_clear}. 
\begin{table*}[t]
\caption{\label{tab:results_clear}%
Simulation results for the alignment degree $P_N$ of three, four, and five $(N=3, 4, 5)$ clusters and experimental data from the Pamir experiment \cite{pamir4}. The size of central cluster $r_{\rm min}=1$ mm, the height $h$ is fixed at 1 km throughout this work. The collision energy in the HYDJET++ generator is 5.02 TeV per nucleon pair, and the generation here and below always includes both components of the model--soft and hard. } 
\begin{ruledtabular}
\begin{tabular}{l c c c}
\setlength{\tabcolsep}{8pt} 
\textbf{Alignment degree} & $\bm{P_3}$ & $\bm{P_4}$ & $\bm{P_5}$ \\
\hline
Pamir results \cite{pamir4} & $0.83\pm 0.27$ & $0.67\pm 0.33$ & $0.33\pm 0.23$ \\
\hline
\diagbox{$r_{\text{res}}$, mm}{centrality} & 
\begin{tabular}{c@{\hspace{0.5em}}c@{\hspace{0.5em}}c} 0\%-5\% & 40\%-75\% & 0\%-75\% \end{tabular} &
\begin{tabular}{c@{\hspace{0.5em}}c@{\hspace{0.5em}}c} 0\%-5\% & 40\%-75\% & 0\%-75\% \end{tabular} &
\begin{tabular}{c@{\hspace{0.5em}}c@{\hspace{0.5em}}c} 0\%-5\% & 40\%-75\% & 0\%-75\% \end{tabular} \\
\hline
$0.5$ & 
\begin{tabular}{c@{\hspace{0.5em}}c@{\hspace{0.5em}}c} 0.23 & 0.27 & 0.26 
\end{tabular} &
\begin{tabular}{c@{\hspace{0.5em}}c@{\hspace{0.5em}}c} 0.032 & 0.061 & 0.053 
\end{tabular} &
\begin{tabular}{c@{\hspace{0.5em}}c@{\hspace{0.5em}}c} 0.005 & 0.015 & 0.011 
\end{tabular} \\
$1$ & 
\begin{tabular}{c@{\hspace{0.5em}}c@{\hspace{0.5em}}c} 0.23 & 0.28 & 0.26 
\end{tabular} &
\begin{tabular}{c@{\hspace{0.5em}}c@{\hspace{0.5em}}c} 0.037 & 0.061 & 0.048 
\end{tabular} &
\begin{tabular}{c@{\hspace{0.5em}}c@{\hspace{0.5em}}c} 0.006 & 0.013 & 0.009 
\end{tabular} \\
$2$ & 
\begin{tabular}{c@{\hspace{0.5em}}c@{\hspace{0.5em}}c} 0.25 & 0.28 & 0.26
\end{tabular} &
\begin{tabular}{c@{\hspace{0.5em}}c@{\hspace{0.5em}}c} 0.042 & 0.058 & 0.048 
\end{tabular} &
\begin{tabular}{c@{\hspace{0.5em}}c@{\hspace{0.5em}}c} 0.007 & 0.013 & 0.009 
\end{tabular} \\
$5$ & 
\begin{tabular}{c@{\hspace{0.5em}}c@{\hspace{0.5em}}c} 0.26 & 0.27 & 0.26
\end{tabular} &
\begin{tabular}{c@{\hspace{0.5em}}c@{\hspace{0.5em}}c} 0.045 & 0.052 & 0.046 
\end{tabular} &
\begin{tabular}{c@{\hspace{0.5em}}c@{\hspace{0.5em}}c} 0.007 & 0.011 & 0.009 
\end{tabular} \\
\end{tabular}
\end{ruledtabular}
\end{table*}

Table \ref{tab:results_clear} clearly shows that the results obtained for the alignment degree $P_N $ of three, four, and five $(N=3, 4, 5)$ clusters are not consistent with the experimental data from the Pamir experiment, and the dependence on cluster size $r_{\rm res}$ is rather weak for all centrality classes $c=\%0-5\%, \%40-75\%, \%0-75\%$. Furthermore, these results are similar to those obtained in previous studies \cite{Lokhtin:2023tze, Lokhtin:2024sbm}, where no clustering procedure for secondary particles was applied. 

The reason for this is straightforward: although angular correlations are present in the HYDJET++ model, including correlations with the reaction plane, they are clearly insufficient to generate the correlations required for the emergence of the alignment phenomenon without a dedicated selection procedure. This emphasizes that the angular distribution plays a more significant role than the radial distribution in the appearance of alignment. Two distinct approaches can be identified. In the geometrical approach~\cite{Lokhtin:2023tze}, the probability density of spots is proportional to $r$, implying a linear increase with radial distance. In the HYDJET++ model, by contrast, the radial distribution follows $F(r) r$, where $F(r)$ is a rapidly decreasing function of $r$. Despite the substantial differences in the radial distributions between these two approaches, the insufficiency of angular correlations in both cases underlines the necessity of the selection procedure for reproducing the observed alignment phenomenon.

Furthermore, it is important to note that the comparison with the Pamir experiment is qualitative in nature, since different types of nuclei are involved (light vs heavy), as well as different kinematic regimes. However, as already mentioned, no data on the manifestation of azimuthal alignment of particles in collider experiments currently exist.

\section{Influence of Transverse Momentum Conservation on Alignment}
\label{sec:pt_conservation}
Since the Pamir Collaboration reported that the alignment phenomenon begins to appear once the detected particle energies exceed a certain threshold, this observation motivated us to attempt to interpret the observed azimuthal correlations in terms of purely kinematic relations, without invoking specific dynamical assumptions or additional mechanisms. We refer to this approach as the transverse momentum disbalance of the most energetic particles and/or clusters, which characterizes the degree to which transverse momentum conservation is preserved in the calculation of alignment.

\subsection{Concept of event-by-event transverse momentum conservation}
As discussed above, the observed alignment may be influenced by purely kinematic effects rather than specific dynamical mechanisms. In particular, in statistical models such as \textsc{HYDJET++}, soft particles are produced independently, leading to event-by-event fluctuations of total momentum, energy, and particle number. These quantities are conserved only on average over many events, typically within a limited rapidity interval, rather than exactly in each individual collision. Consequently, the total transverse momentum of all particles vanishes statistically--but not precisely--in a single event, which can affect azimuthal correlations among the most energetic particles.
To account for this effect, we introduce after performing the clustering procedure an event-level constraint on the residual transverse momentum,
\begin{equation}
\label{a}
\left|{\bf p}_{T_1} + {\bf p}_{T_2} + \dots + {\bf p}_{T_{N-1}}\right| < \Delta,
\end{equation}
where ${\bf p}_{T_i}$ is the transverse momentum of the $i$th cluster. The parameter $\Delta$ defines the allowed degree of transverse momentum disbalance: smaller $\Delta$ values correspond to stronger event-by-event momentum conservation, while larger ones permit greater deviations. It is important to note that, in applying this approach to the HYDJET++ model, we do not modify its parameters or introduce any changes to the generator. Rather, we work with the particles after generation, i.e., we first select those that satisfy the clustering condition~\eqref{clust_cond}, and subsequently the resulting clusters are tested against condition~\eqref{a}.

The alignment degree $P_N$, defined in Eq.~\eqref{deg_align}, is evaluated as a function of the transverse momentum disbalance parameter~$\Delta$,
\begin{equation}
\label{Pnd}
P_{N}(\Delta) = \frac{l^{[\Delta]}}{L^{[\Delta]}},
\end{equation}
where $l^{[\Delta]}$ denotes the number of events with $\lambda_N > 0.8$, and $L^{[\Delta]}$ is the number of events satisfying condition~\eqref{a}. This formulation allows us to quantify how the alignment degree $P_N(\Delta)$ depends on the level of transverse momentum conservation for the $N = 3, 4, 5$ most energetic particles and clusters on an event-by-event basis within a limited rapidity range. The total number of generated events used in the analysis is fixed at $L^{[\mathrm{tot}]} = (1\text{--}3)\cdot 10^6$ (see Appendix~\ref{Stats0-5}, \ref{Stats40-75}, and~\ref{Stats0-75} for details).

\subsection{Simulation results with local $p_T$ conservation}
After introducing the basic definitions and the transverse momentum disbalance \eqref{a} for clusters of secondary particles, we now present the simulation results for the alignment degree $P_N$ as a function of this disbalance $\Delta$. As discussed in the previous Sections, the results are shown for three centrality classes of heavy-ion collisions: $0\%-5 \%$, $40\%-75\%$, and $0\%-75\%$.

\begin{figure*}[!t]
\begin{center}
\includegraphics[scale=0.55]{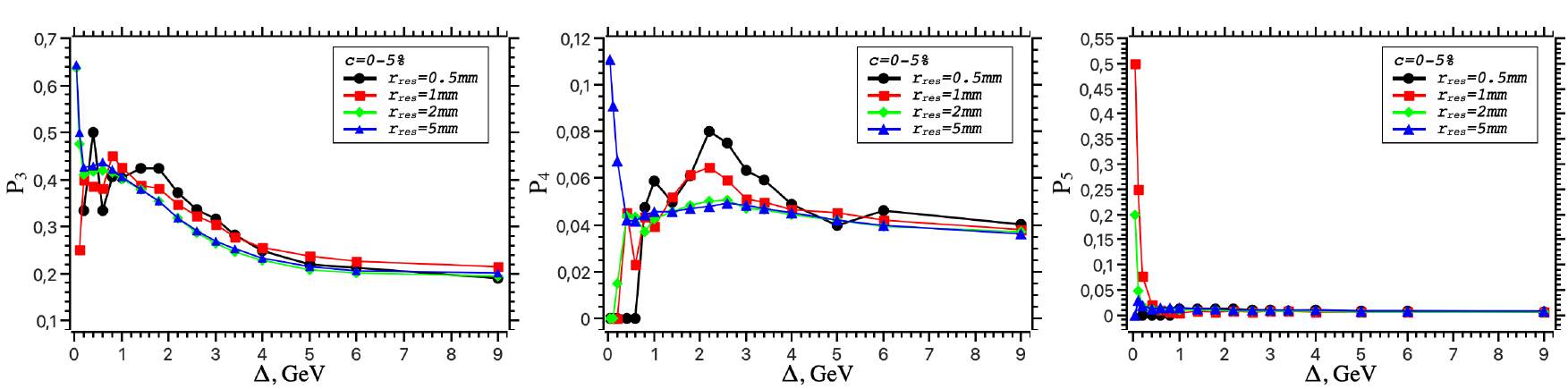}
\caption{Degree of alignment $P_3, P_4, P_5$ for the three, four, five clusters as a function of the disbalance $\Delta$ at the different values of the resolution parameter $r_{\rm res}=0.5,1,2,5$ mm. Centrality class $c=0\%-5 \%$.}
\label{fig:PN_0-5}
\end{center}
\end{figure*}

\begin{figure*}[!t]
\begin{center}
\includegraphics[scale=0.55]{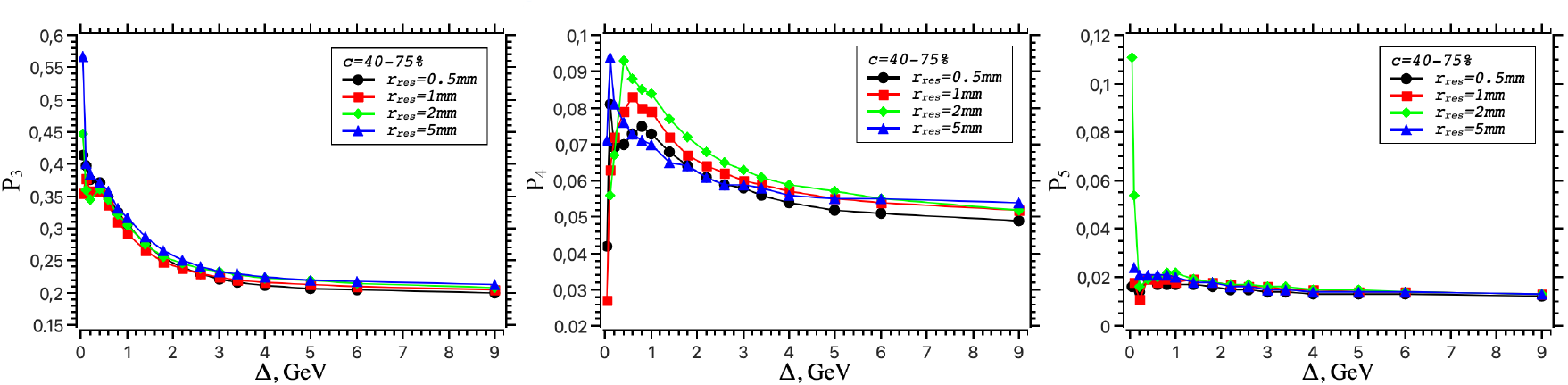}
\caption{Degree of alignment $P_3, P_4, P_5$ for the three, four, five clusters as a function of the disbalance $\Delta$ at the different values of the resolution parameter $r_{\rm res}=0.5,1,2,5$ mm. Centrality class $c=40\%-75 \%$.}
\label{fig:PN_40-75}
\end{center}
\end{figure*}

\begin{figure*}[!t]
\begin{center}
\includegraphics[scale=0.55]{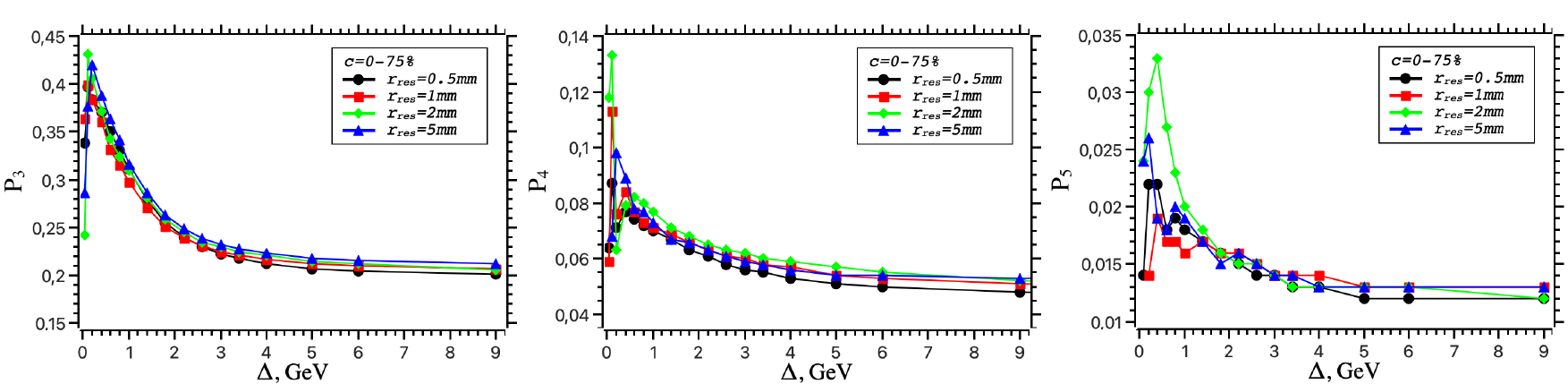}
\caption{Degree of alignment $P_3, P_4, P_5$ for the three, four, five clusters as a function of the disbalance $\Delta$ at the different values of the resolution parameter $r_{\rm res}=0.5,1,2,5$ mm. Centrality class $c=0\%-75 \%$.}
\label{fig:PN_0-75}
\end{center}
\end{figure*}

Figures~\ref{fig:PN_0-5}, \ref{fig:PN_40-75}, and \ref{fig:PN_0-75} present the dependence of the alignment degree $P_N$ on the transverse momentum disbalance $\Delta$ for different values of the cluster size parameter $r_{\mathrm{res}}$ but $r_{\rm min}$ unchanged. Figure~\ref{fig:PN_0-5} corresponds to the most central collisions (0\%--5\%), Fig.~\ref{fig:PN_40-75} to peripheral collisions (40\%--75\%), and Fig.~\ref{fig:PN_0-75} to the combined sample (0\%--75\%). For smaller clusters (lower $r_{\mathrm{res}}$), the alignment degree shows a weaker dependence on $\Delta$, whereas for larger clusters the variation of $P_N$ with disbalance becomes more pronounced. This trend reflects the increasing role of collective effects and geometric asymmetries as the effective cluster size grows.

These results indicate that the alignment effect is sensitive both to the event geometry via centrality class and to the scale at which clusters of secondary particles are formed. Taking into account the transverse momentum disbalance in the conservation of clusters momenta \eqref{a} leads to an increase in the alignment compared to the case without such consideration, as summarized in the Table \ref{tab:results_clear}. Overall, for all centrality classes and cluster multiplicities $N$, our hypothesis is confirmed: the smaller the $\Delta$, the stronger the azimuthal correlation of the clusters, whereas a larger disbalance corresponds to a weaker alignment. The observed dependence of $P_N$ on missing transverse momentum and the cluster size parameter $r_{\mathrm{res}}$ suggests that the interplay between local momentum correlations and global event geometry may play a significant role in shaping the alignment patterns.

\subsection{Results with soft and hard components of HYDJET++ and stable resonances}
The HYDJET++ model makes it possible to simulate heavy-ion collisions with only the soft component, only hard jets, or both components simultaneously, as well as to switch resonance decays on or off. This makes it possible to isolate the role of individual model ingredients in the formation of the alignment effect. In our previous study~\cite{Lokhtin:2024sbm}, the influence of jets and the soft component on alignment was investigated separately; however, clustering of secondary particles was not included. In the present analysis, we extend this study by considering the effect of different HYDJET++ components together with the clustering procedure.
\begin{figure*}[!t]
\begin{center}
\includegraphics[scale=0.55]{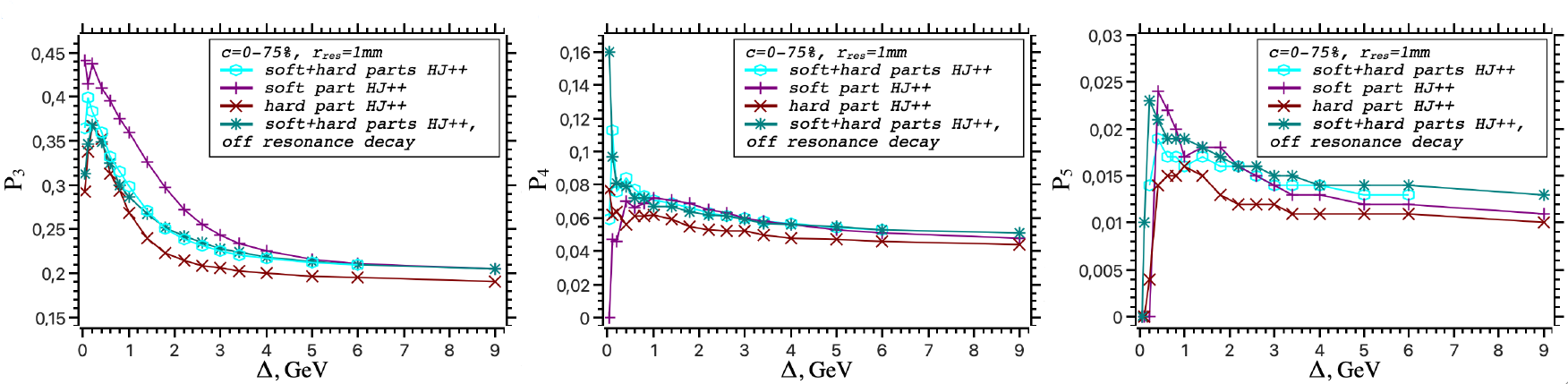}
\caption{Comparison of the alignment degrees $P_3$, $P_4$, and $P_5$ as functions of the transverse momentum disbalance $\Delta$ for the resolution parameter $r_{\rm res}=1$~mm and the centrality class $c=0\%$--$75\%$. The results are shown for events generated with only the soft component, only the hard component, and with both components while resonance decays are switched off.}
\label{fig:COMP_S_H_NR}
\end{center}
\end{figure*}

Figure~\ref{fig:COMP_S_H_NR} presents the alignment degrees $P_3$, $P_4$, and $P_5$ as functions of the transverse momentum disbalance $\Delta$ for the 0\%--75\% centrality class and the resolution parameter $r_{\mathrm{res}} = 1$~mm. The results are shown for events generated with only the soft component, only the hard component, and with both components and resonance decays switched off. For the soft component, the alignment effect appears somewhat stronger than for the hard component, although the difference is quite moderate and noticeable mainly for $P_3$. In this case, it seems easier to achieve the required momentum disbalance due to the softer spectrum and potentially anisotropic flow. A noticeable enhancement of the alignment is also observed for $P_4$ at small $\Delta$ in the absence of resonance decays. In all other cases, no significant influence of individual model components on the alignment is seen. Qualitatively similar behavior was previously obtained without applying clustering~\cite{Lokhtin:2024sbm}.

\section{Comparison of Simulations With and Without Clustering Under Transverse Momentum Conservation}
\label{sec:comparison}
\begin{figure*}[!t]
\begin{center}
\includegraphics[scale=0.55]{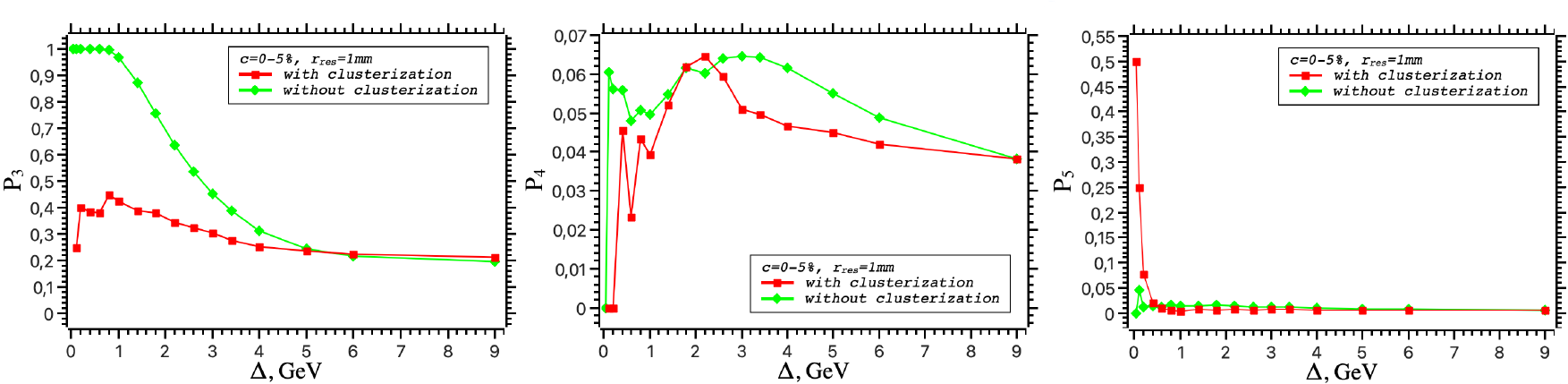}
\caption{Comparison of alignment degree $P_3, P_4, P_5$ as a function of the disbalance $\Delta$ with and without clustering. The resolution parameter $r_{\rm res} = 1 $ mm. Centrality class $c=0\%-5 \%$.}
\label{fig:PN_0-5_comp}
\end{center}
\end{figure*}

\begin{figure*}[!t]
\begin{center}
\includegraphics[scale=0.55]{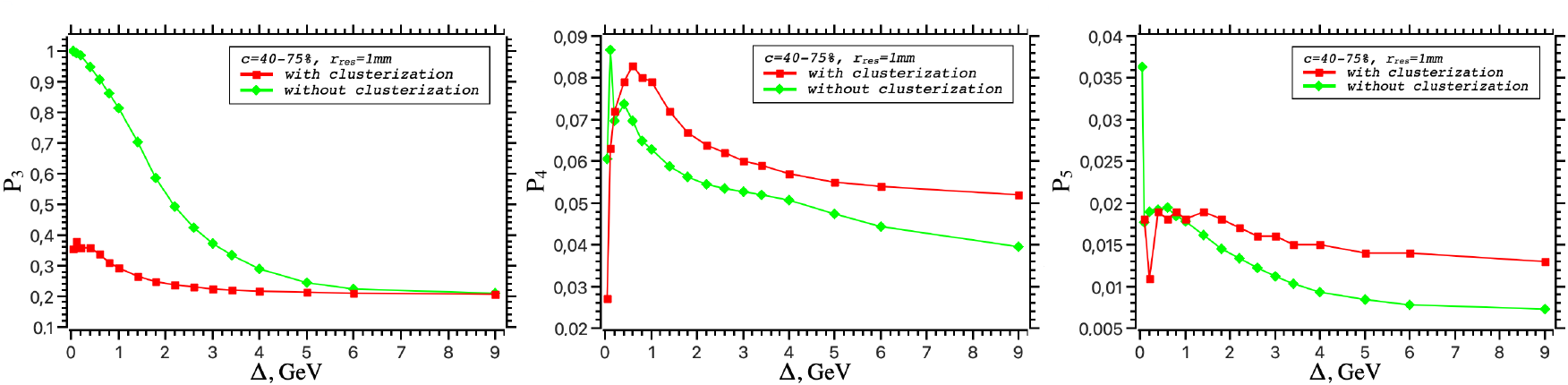}
\caption{Comparison of alignment degree $P_3, P_4, P_5$ as a function of the disbalance $\Delta$ with and without clustering. The resolution parameter $r_{\rm res} = 1 $ mm. Centrality class $c=40\%-75 \%$.}
\label{fig:PN_40-75_comp}
\end{center}
\end{figure*}

\begin{figure*}[!t]
\begin{center}
\includegraphics[scale=0.55]{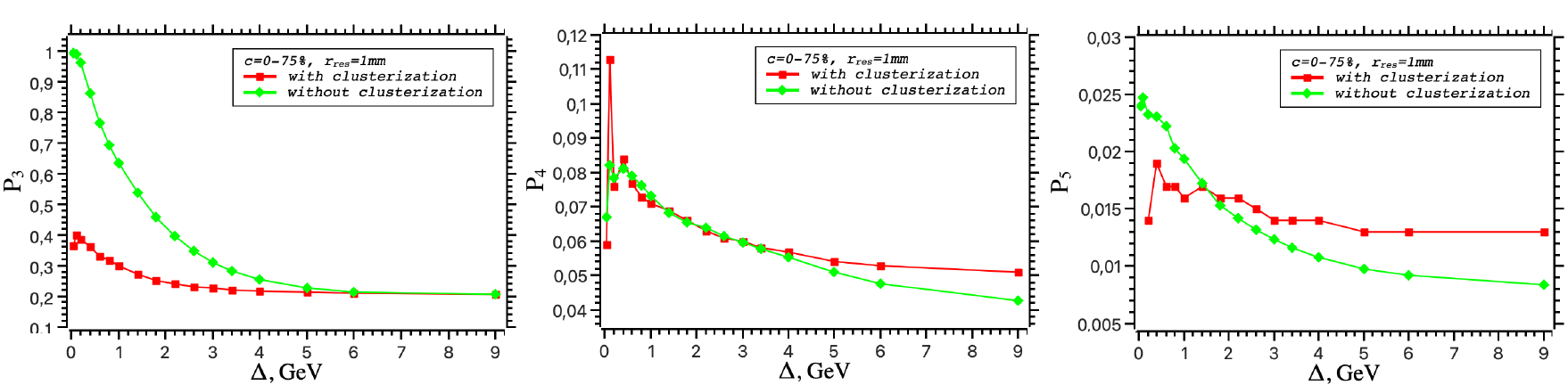}
\caption{Comparison of alignment degree $P_3, P_4, P_5$ as a function of the disbalance $\Delta$ with and without clustering. The resolution parameter $r_{\rm res} = 1 $ mm. Centrality class $c=0\%-75 \%$.}
\label{fig:PN_0-75_comp}
\end{center}
\end{figure*}

Since the main motivation and goal of this work is to demonstrate the impact of secondary particle clustering on the alignment phenomenon, it is both appropriate and logical to devote a separate section to this comparison. The simulation results without clustering can be found in our recent work~\cite{Lokhtin:2024sbm}. Figures~\ref{fig:PN_0-5_comp}, \ref{fig:PN_40-75_comp}, and \ref{fig:PN_0-75_comp} compare the alignment degree simulations with and without clustering for the three centrality classes at a cluster size of $r_{\mathrm{res}} = 1$~mm. 

It is evident that, for all three centrality classes, clustering reduces $P_3$ in a range $\Delta=$=0-1 GeV from its maximum value approximately by a factor of 2. This effect can be attributed to the fact that the direction of the cluster's transverse momentum and its radial position do not necessarily coincide. The cluster momentum is the sum of the momenta of all constituent particles, whereas its position is determined by the energy-weighted average of the particles within it. This discrepancy can lead to a decrease in the observed alignment compared to the case without clustering.

In the case of four clusters, $P_4$ , the situation is somewhat different: for the most central collisions ($c=0\%-5 \%$), clustering reduces the alignment over a wide range of the transverse momentum disbalance $\Delta$, whereas the inclusion of peripheral collisions generally increases the alignment, which may reflect the influence of anisotropic azimuthal flow of secondary particles.

For five clusters, $P_5$, the results are more unexpected: central collisions exhibit a significant increase in alignment with clustering compared to the case without clustering at small $\Delta$ values, although this result has relatively low statistics (see Appendix \ref{Stats0-5}). Peripheral collisions ($c=40\%-75 \%$ and $c=0\%-75 \%$), on the other hand, clearly enhance the alignment relative to the nonclustered case in the range $\Delta > 1$~GeV, which may also be a consequence of the anisotropy in the secondary particle flow.

Overall, by examining Figs.~\ref{fig:PN_0-5_comp} -- \ref{fig:PN_0-75_comp}, we can conclude that the effect of secondary particle clustering on the alignment phenomenon is indeed present. Its manifestation is not uniform and depends on the collision centrality, the number of energetically distinguished clusters under study $N$, as well as on the event-by-event conservation of the cluster transverse momenta.

\section{Discussion and Conclusions}
\label{sec:conclusion}
In this work, we present a broader perspective on the alignment phenomenon, extending our general view formulated in Refs.~\cite{Lokhtin:2023tze,Lokhtin:2024trs} and further developing the previously proposed approach to modeling azimuthal alignment in heavy-ion collisions~\cite{Lokhtin:2024sbm} using the HYDJET++ event generator. We demonstrate that the combined effect of particle clustering and event-by-event transverse momentum conservation can significantly affect the observed alignment. Clusters are constructed through an iterative pairwise comparison of particle coordinates on the emulsion ``film,'' with the cluster position being updated each time a new particle is added, until no pair satisfies the distance condition $d_{ij} > r_\text{res}$ (see Sec.~\ref{sec:simulation} for details).

The simulations were performed for three centrality classes: $0\%-5 \%$, $40\%-75\%$, and $0\%-75\%$. These correspond, respectively, to nearly central collisions with maximal clustering activity, peripheral collisions dominated by anisotropic flow and minimal clustering, and a more realistic mixed scenario that includes all collision types. The simulations of Pb+Pb collisions at $\sqrt{s}=5.02$ TeV per nucleon pair include both the soft and hard components of the HYDJET++ model, with the height of the primary interaction fixed at $h = 1$~km and the central cluster size set to $r_\text{res} = 1$~mm. It is important to note that only a qualitative and, to some extent, forced comparison with the Pamir experiment can be made, since no evidence of the alignment effect has been reported in collider experiments.

\begin{table*}[!t]
\caption{Comparison of our best simulation results for alignment with the Pamir experimental data.}
\label{tab:final_results}
\begin{ruledtabular}
\begin{tabular}{cccc}
\textbf{Alignment degree} & $\bm{P_3}$ & $\bm{P_4}$ & $\bm{P_5}$ \\
\hline
Pamir results \cite{pamir4} & $0.83\pm 0.27$ & $0.67\pm 0.33$ & $0.33\pm 0.23$ \\
Our results with clustering & 0.65 & 0.13 & 0.5  \\
Our results without clustering \cite{Lokhtin:2024sbm}  & 1.0 & 0.07 & 0.04 \\
Reference \cite{Lokhtin:2023tze} & 0.2 & 0.04 & 0.008 \\
\end{tabular}
\end{ruledtabular}
\end{table*}

Table~\ref{tab:final_results} shows a comparison of our best simulation results with the Pamir experimental data for three-, four-, and five-cluster configurations. The ``Reference'' row corresponds to the alignment degree values for an isotropic particle distribution, consistent with the results of Ref.~\cite{Lokhtin:2023tze}, which provides a purely geometrical interpretation of the alignment phenomenon, and with the recent study~\cite{Lokhtin:2024sbm} employing the HYDJET++ event generator. These findings are also in agreement with the results obtained in the present work (see Table~\ref{tab:results_clear}). Overall, the comparison demonstrates a satisfactory correspondence between the simulated and experimental values, indicating that the implemented model captures the essential features of the observed alignment effect.

The clustering procedure itself does not lead to a significant enhancement of the alignment degree, as can also be seen from Table~\ref{tab:results_clear}. Therefore, a natural next step was to apply the approach developed in our previous works~\cite{Lokhtin:2023tze, Lokhtin:2024sbm}, which accounts for the missing total transverse momentum of clusters, or the transverse momentum disbalance, i.e., event-by-event transverse momentum conservation~\eqref{a}. This method results in a considerable increase in the simulated alignment (Table~\ref{tab:final_results}) and, at the same time, does not require introducing or modeling any complex dynamical processes in relativistic nucleus–nucleus collisions.

As seen from Figs.~\ref{fig:PN_0-5}--\ref{fig:PN_0-75}, the concept of the missing transverse momentum, combined with the selection of the most energetic clusters, provides the highest values of the alignment degree in the disbalance range $\Delta = 0 \text{--} 1$~GeV. This confirms our hypothesis that minimizing the total transverse momentum of clusters directly affects their azimuthal alignment. Moreover, there is a clear tendency that increasing the cluster size $r_{\mathrm{res}}$ leads to a stronger response in the region $\Delta = 0 \text{--}1 $~GeV for all centrality classes and cluster multiplicities $N$.

A comparison of events generated with only the soft component, only the hard component, and with both components and resonance decays switched off (Fig.~\ref{fig:COMP_S_H_NR}) suggests that the alignment pattern shows only a moderate sensitivity to the specific particle production mechanism. The alignment is somewhat stronger for the soft component, mainly for $P_3$, which may be related to the softer momentum spectrum and possible anisotropic flow facilitating the required transverse momentum disbalance. A modest enhancement of the alignment for $P_4$ at small $\Delta$ appears only when resonance decays are excluded, while in all other cases the influence of individual model ingredients remains limited. This behavior is consistent with previous observations obtained without applying clustering~\cite{Lokhtin:2024sbm} and indicates that the overall alignment signal remains relatively robust with respect to variations in the underlying production dynamics, with resonance decays playing only a minor role in modifying the signal.

In Figs.~\ref{fig:PN_0-5_comp}--\ref{fig:PN_0-75_comp} we present the influence of the clustering procedure at a fixed resolution radius $r_{\rm res} = 1$~mm. For three clusters ($N=3$), the alignment parameter decreases substantially in all centrality classes; this effect is attributed to differing directions of the transverse momenta and the radial positions of the clusters (see Sec.~\ref{sec:comparison} for details). For four clusters ($N=4$), the impact of clustering becomes especially noticeable when combined with the anisotropic flow of secondary particles, i.e., for centralities of $40\%-75\%$ and $0\%-75\%$. For five clusters ($N=5$), the case of the most central collisions ($0-5\%$) stands out: here clustering strongly enhances the alignment at small $\Delta$; however, this result is based on limited statistics (only a few events, see Appendix \ref{Stats0-5}). When peripheral collisions are also included, the alignment increases for $\Delta > 1$~GeV.

Referring to Table~\ref{tab:final_results}, which presents a comparison between our best simulation results for the alignment effect and the data from the Pamir Collaboration, we can see that, in general, our concept provides a reasonable description of the experimental observations. However, we emphasize once again that this comparison should be regarded as mostly qualitative and, to some extent, forced, since no data on the observation of azimuthal alignment of particles at collider experiments are currently available.

Distinctive features of our approach include the type of colliding nuclei, in our case heavy lead nuclei rather than lighter ones such as iron or oxygen. Nevertheless, there exist studies indicating a similarity of the observed effects in systems with both high and low nucleon densities~\cite{Pasechnik:2016wkt,PHENIX:2018lia,Sahoo:2023tmn}. Moreover, in Ref.~\cite{Lokhtin:2005bb} indications of a significant alignment of particles via jets mechanism production with respect to reference values were obtained in proton-proton collisions at $\sqrt{s}=14$~TeV using the PYTHIA event generator~\cite{Sjostrand:2000wi}.

The alignment is sensitive to nonflow effects and reflects the relative arrangement of nearby particle clusters, providing a direct probe of local structures in the azimuthal plane. Its limitations are related to reduced statistical stability -- i.e., the effect has not been observed in collider experiments within the accessible rapidity range -- as well as the lack of a direct connection to the azimuthal collective flow. In contrast, the Fourier decomposition, namely $v_2$ and the correlator $c_2\{2\}$, are statistically more robust observables directly connected to the collective dynamics of particles, although they may mix flow and nonflow contributions to the anisotropy.

Undoubtedly, we cannot assert with confidence that the alignment phenomenon arises solely from the selection of the most energetic clusters and clusterlike particles together with the conservation of their total transverse momentum. Nevertheless, the results obtained in this work do not exclude the possibility that additional, possibly novel, mechanisms may contribute to the alignment patterns reported in cosmic-ray experiments, thereby motivating dedicated searches for similar types of azimuthal correlations (under appropriate kinematic conditions) in collider environments. We therefore leave room for further discussion and investigation, as the observed alignment may also be shaped by complex dynamical processes in the quark-gluon plasma, collective effects, and phenomena beyond.

\acknowledgments
We are grateful to A.S.~Chernyshov and A.I.~Demianov for valuable and stimulating discussions, and to S.N.~Nedelko for helpful comments and communication. This work was supported by the Russian Science Foundation under Grant No.~24-22-00011.

\appendix

\section{Statistics of modeling}

\clearpage
\subsection{Centrality 0\%-5\%}
\label{Stats0-5}

\begin{table*}[!b]
\centering
\caption{Statistics of the alignment degree $P_3$ versus disbalance $\Delta$ using clustering with different cluster sizes $r_{\rm res}$. Centrality class $c=0-5\%$. The alignment degree  $P_3(\Delta)=l^{[\Delta]}/L^{[\Delta]}$, and $L^{[\rm tot]} = 10^6$ denotes the total number of simulated events.}
\label{tab:STAT_3P_0-5}
\begin{tabular}{c|ccc@{\hskip 8pt}|ccc@{\hskip 8pt}|ccc@{\hskip 8pt}|ccc@{\hskip 8pt}}
\toprule
\multirow{2}{*}{$\Delta$,~GeV} 
 & \multicolumn{3}{c}{$r_{\rm res}=0.5~\rm mm$} 
 & \multicolumn{3}{c}{$r_{\rm res}=1~\rm mm$} 
 & \multicolumn{3}{c}{$r_{\rm res}=2~\rm mm$} 
 & \multicolumn{3}{c}{$r_{\rm res}=5~\rm mm$} \\
\cmidrule(lr){2-4} \cmidrule(lr){5-7} \cmidrule(lr){8-10} \cmidrule(lr){11-13}
 & $P_3$ & $l^{[\Delta]}$ & $L^{[\Delta]}$  
 & $P_3$ & $l^{[\Delta]}$ & $L^{[\Delta]}$ 
 & $P_3$ & $l^{[\Delta]}$ & $L^{[\Delta]}$ 
 & $P_3$ & $l^{[\Delta]}$ & $L^{[\Delta]}$ \\
\midrule
0    & --      & 0      & 0      & --      & 0      & 0      & --      & 0      & 0      & --      & 0      & 0      \\
0.05   & --      & 0      & 0      & 0   & 0      & 2      & 0.636   & 7      & 11     & 0.643   & 9      & 14     \\
0.1    & --      & 0      & 0      & 0.250   & 1      & 4      & 0.476   & 10     & 21     & 0.500   & 19     & 38     \\
0.2    & 0.333   & 1      & 3      & 0.400   & 2      & 5      & 0.410   & 48     & 117    & 0.426   & 72     & 169    \\
0.4    & 0.500   & 5      & 10     & 0.385   & 10     & 26     & 0.416   & 179    & 430    & 0.427   & 280    & 655    \\
0.6    & 0.333   & 8      & 24     & 0.381   & 24     & 63     & 0.419   & 335    & 799    & 0.437   & 595    & 1363   \\
0.8    & 0.405   & 17     & 42     & 0.450   & 49     & 109    & 0.415   & 530    & 1276   & 0.422   & 906    & 2147   \\
1.0    & 0.404   & 23     & 57     & 0.426   & 66     & 155    & 0.400   & 699    & 1747   & 0.406   & 1226   & 3017   \\
1.4    & 0.423   & 33     & 78     & 0.387   & 98     & 253    & 0.378   & 1059   & 2801   & 0.379   & 1798   & 4744   \\
1.8    & 0.423   & 41     & 97     & 0.380   & 124    & 326    & 0.353   & 1326   & 3754   & 0.354   & 2255   & 6378   \\
2.2    & 0.372   & 45     & 121    & 0.346   & 142    & 410    & 0.318   & 1489   & 4689   & 0.318   & 2525   & 7947   \\
2.6    & 0.336   & 52     & 155    & 0.323   & 159    & 492    & 0.286   & 1588   & 5555   & 0.291   & 2719   & 9354   \\
3.0    & 0.315   & 57     & 181    & 0.304   & 170    & 559    & 0.264   & 1676   & 6340   & 0.269   & 2872   & 10659  \\
3.4    & 0.282   & 57     & 202    & 0.279   & 178    & 639    & 0.246   & 1735   & 7063   & 0.252   & 2981   & 11808  \\
4.0    & 0.249   & 59     & 237    & 0.255   & 185    & 726    & 0.228   & 1816   & 7975   & 0.234   & 3116   & 13329  \\
5.0    & 0.219   & 62     & 283    & 0.236   & 202    & 855    & 0.209   & 1939   & 9272   & 0.214   & 3301   & 15429  \\
6.0    & 0.212   & 66     & 312    & 0.226   & 217    & 961    & 0.202   & 2036   & 10091  & 0.206   & 3463   & 16825  \\
9.0    & 0.190   & 77     & 405    & 0.215   & 241    & 1123   & 0.195   & 2158   & 11043  & 0.201   & 3681   & 18298  \\
12.0   & 0.187   & 88     & 471    & 0.211   & 255    & 1209   & 0.195   & 2196   & 11244  & 0.200   & 3717   & 18569  \\
15.0   & 0.191   & 102    & 534    & 0.210   & 267    & 1273   & 0.195   & 2216   & 11342  & 0.200   & 3731   & 18651  \\
18.0   & 0.190   & 112    & 590    & 0.207   & 275    & 1328   & 0.195   & 2224   & 11395  & 0.200   & 3737   & 18685  \\
\bottomrule
\end{tabular}
\end{table*}

\clearpage

\begin{table*}[!b]
\centering
\caption{Statistics of the alignment degree $P_4$ versus disbalance $\Delta$ using clustering with different cluster sizes $r_{\rm res}$. Centrality class $c=0-5\%$. The alignment degree  $P_4(\Delta)=l^{[\Delta]}/L^{[\Delta]}$, and $L^{[\rm tot]} = 10^6$ denotes the total number of simulated events.}
\label{tab:STAT_4P_0-5}
\begin{tabular}{c|ccc@{\hskip 8pt}|ccc@{\hskip 8pt}|ccc@{\hskip 8pt}|ccc}
\toprule
\multirow{2}{*}{$\Delta$,~GeV} 
 & \multicolumn{3}{c}{$r_{\rm res}=0.5~\rm mm$} 
 & \multicolumn{3}{c}{$r_{\rm res}=1~\rm mm$} 
 & \multicolumn{3}{c}{$r_{\rm res}=2~\rm mm$} 
 & \multicolumn{3}{c}{$r_{\rm res}=5~\rm mm$} \\
\cmidrule(lr){2-4} \cmidrule(lr){5-7} \cmidrule(lr){8-10} \cmidrule(lr){11-13}
 & $P_4$ & $l^{[\Delta]}$ & $L^{[\Delta]}$ 
 & $P_4$ & $l^{[\Delta]}$ & $L^{[\Delta]}$ 
 & $P_4$ & $l^{[\Delta]}$ & $L^{[\Delta]}$ 
 & $P_4$ & $l^{[\Delta]}$ & $L^{[\Delta]}$ \\
\midrule
0  & --      & 0   & 0     & 0       & 0   & 0     & 0       & 0   & 0     & 0       & 0   & 0     \\
0.05   & 0       & 0   & 2     & 0       & 0   & 0     & 0       & 0   & 6     & 0.111   & 1   & 9     \\
0.1    & 0    & 0   & 3     & 0    & 0   & 2     & 0    & 0   & 17    & 0.091   & 3   & 33    \\
0.2    & 0    & 0   & 4     & 0   & 0   & 5     & 0.015   & 1   & 67    & 0.068   & 9   & 133   \\
0.4    & 0    & 0   & 8     & 0.045   & 1   & 22    & 0.044   & 11  & 251   & 0.042   & 19  & 453   \\
0.6    & 0    & 0   & 14    & 0.023   & 1   & 43    & 0.044   & 23  & 527   & 0.042   & 39  & 939   \\
0.8    & 0.048   & 1   & 21    & 0.043   & 3   & 69    & 0.037   & 34  & 917   & 0.044   & 70  & 1576  \\
1.0    & 0.059   & 2   & 34    & 0.039   & 4   & 102   & 0.042   & 56  & 1322  & 0.046   & 104 & 2285  \\
1.4    & 0.050   & 3   & 60    & 0.052   & 9   & 173   & 0.046   & 100 & 2182  & 0.046   & 174 & 3808  \\
1.8    & 0.061   & 5   & 82    & 0.062   & 16  & 259   & 0.049   & 148 & 3046  & 0.047   & 247 & 5243  \\
2.2    & 0.080   & 9   & 112   & 0.065   & 21  & 325   & 0.050   & 198 & 3928  & 0.048   & 317 & 6600  \\
2.6    & 0.075   & 10  & 133   & 0.059   & 24  & 404   & 0.051   & 242 & 4788  & 0.049   & 391 & 7922  \\
3.0    & 0.063   & 10  & 158   & 0.051   & 24  & 470   & 0.047   & 264 & 5596  & 0.048   & 445 & 9204  \\
3.4    & 0.059   & 11  & 185   & 0.050   & 27  & 544   & 0.046   & 293 & 6313  & 0.047   & 488 & 10403 \\
4.0    & 0.049   & 11  & 225   & 0.047   & 30  & 642   & 0.045   & 323 & 7256  & 0.045   & 542 & 11969 \\
5.0    & 0.040   & 11  & 275   & 0.045   & 35  & 776   & 0.042   & 355 & 8473  & 0.042   & 598 & 14155 \\
6.0    & 0.046   & 15  & 325   & 0.042   & 37  & 883   & 0.039   & 370 & 9418  & 0.040   & 629 & 15811 \\
9.0    & 0.040   & 17  & 422   & 0.038   & 42  & 1102  & 0.037   & 402 & 10889 & 0.036   & 663 & 18223 \\
12.0   & 0.040   & 19  & 479   & 0.041   & 49  & 1202  & 0.037   & 413 & 11253 & 0.036   & 674 & 18776 \\
15.0   & 0.041   & 22  & 537   & 0.039   & 50  & 1266  & 0.037   & 424 & 11378 & 0.036   & 682 & 18908 \\
18.0   & 0.039   & 23  & 584   & 0.040   & 53  & 1313  & 0.037   & 429 & 11444 & 0.036   & 685 & 18963 \\
\bottomrule
\end{tabular}
\end{table*}


\begin{table*}
\centering
\caption{Statistics of the alignment degree $P_5$ versus disbalance $\Delta$ using clustering with different cluster sizes $r_{\rm res}$. Centrality class $c=0-5\%$. The alignment degree  $P_5(\Delta)=l^{[\Delta]}/L^{[\Delta]}$, and $L^{[\rm tot]}=3 \cdot 10^6$ denotes the total number of simulated events.}
\label{tab:STAT_5P_0-5}
\begin{tabular}{c|ccc@{\hskip 8pt}|ccc@{\hskip 8pt}|ccc@{\hskip 8pt}|ccc@{\hskip 8pt}}
\toprule
\multirow{2}{*}{$\Delta$,~GeV} 
 & \multicolumn{3}{c}{$r_{\rm res}=0.5~\rm mm$} 
 & \multicolumn{3}{c}{$r_{\rm res}=1~\rm mm$} 
 & \multicolumn{3}{c}{$r_{\rm res}=2~\rm mm$} 
 & \multicolumn{3}{c}{$r_{\rm res}=5~\rm mm$} \\
\cmidrule(lr){2-4} \cmidrule(lr){5-7} \cmidrule(lr){8-10} \cmidrule(lr){11-13}
 & $P_5$ & $l^{[\Delta]}$ & $L^{[\Delta]}$ 
 & $P_5$ & $l^{[\Delta]}$ & $L^{[\Delta]}$ 
 & $P_5$ & $l^{[\Delta]}$ & $L^{[\Delta]}$ 
 & $P_5$ & $l^{[\Delta]}$ & $L^{[\Delta]}$ \\
\midrule
0    & --      & 0    & 0     & --      & 0    & 0     & --      & 0    & 0     & --      & 0    & 0     \\
0.05 & --      & 0    & 0     & 0.500    & 1    & 2     & 0.200    & 2    & 10    & 0       & 0    & 20    \\
0.1  & 0    & 0    & 0     & 0.250    & 1    & 4     & 0.048   & 2    & 41    & 0.029   & 2    & 67    \\
0.2  & 0    & 0    & 3     & 0.077   & 1    & 13    & 0.018   & 3    & 171   & 0.018   & 5    & 285   \\
0.4  & 0    & 0    & 12    & 0.020   & 1    & 51    & 0.013   & 8    & 630   & 0.013   & 14   & 1084  \\
0.6  & 0    & 0    & 31    & 0.010   & 1    & 105   & 0.012   & 17   & 1379  & 0.015   & 35   & 2319  \\
0.8  & 0    & 0    & 56    & 0.006   & 1    & 179   & 0.011   & 27   & 2358  & 0.015   & 57   & 3904  \\
1.0     & 0.011   & 1    & 90    & 0.004   & 1    & 268   & 0.010   & 35   & 3474  & 0.013   & 76   & 5687  \\
1.4  & 0.012   & 2    & 169   & 0.008   & 4    & 472   & 0.010   & 60   & 5877  & 0.012   & 118  & 9752  \\
1.8  & 0.012   & 3    & 243   & 0.006   & 4    & 709   & 0.010   & 82   & 8485  & 0.011   & 161  & 14053 \\
2.2  & 0.013   & 4    & 318   & 0.008   & 7    & 914   & 0.009   & 97   & 11019 & 0.011   & 192  & 18245 \\
2.6  & 0.010   & 4    & 418   & 0.006   & 7    & 1103  & 0.008   & 110  & 13391 & 0.010   & 232  & 22296 \\
3.0     & 0.010   & 5    & 493   & 0.007   & 9    & 1315  & 0.008   & 123  & 15598 & 0.010   & 260  & 26075 \\
3.4  & 0.009   & 5    & 566   & 0.007   & 11   & 1533  & 0.007   & 133  & 17745 & 0.010   & 285  & 29695 \\
4.0     & 0.009   & 6    & 661   & 0.006   & 11   & 1808  & 0.007   & 149  & 20654 & 0.009   & 311  & 34584 \\
5.0     & 0.007   & 6    & 841   & 0.006   & 14   & 2249  & 0.007   & 163  & 24856 & 0.008   & 340  & 41621 \\
6.0     & 0.007   & 7    & 984   & 0.006   & 15   & 2578  & 0.006   & 171  & 28220 & 0.008   & 363  & 47052 \\
9.0     & 0.006   & 8    & 1290  & 0.006   & 20   & 3320  & 0.006   & 196  & 33898 & 0.007   & 393  & 56165 \\
12.0    & 0.006   & 9    & 1505  & 0.006   & 21   & 3718  & 0.006   & 202  & 35684 & 0.007   & 399  & 59098 \\
15.0   & 0.005   & 9    & 1706  & 0.006   & 22   & 3955  & 0.006   & 202  & 36209 & 0.007   & 403  & 59794 \\
18.0   & 0.006   & 11   & 1883  & 0.006   & 25   & 4144  & 0.006   & 203  & 36438 & 0.007   & 404  & 60018 \\
\bottomrule
\end{tabular}
\end{table*}

\clearpage
\newpage
\subsection{Centrality 40\%-75\%}
\label{Stats40-75}

\begin{table*}[!b]
\centering
\caption{Statistics of the alignment degree $P_3$ versus disbalance $\Delta$ using clustering with different cluster sizes $r_{\rm res}$. Centrality class $c=40-75\%$. The alignment degree  $P_3(\Delta)=l^{[\Delta]}/L^{[\Delta]}$, and $L^{[\rm tot]} = 10^6$ denotes the total number of simulated events.}
\label{tab:STAT_3P_40-75}
\begin{tabular}{c|ccc@{\hskip 8pt}|ccc@{\hskip 8pt}|ccc@{\hskip 8pt}|ccc@{\hskip 8pt}}
\toprule
\multirow{2}{*}{$\Delta$,~GeV} 
 & \multicolumn{3}{c}{$r_{\rm res}=0.5~\rm mm$} 
 & \multicolumn{3}{c}{$r_{\rm res}=1~\rm mm$} 
 & \multicolumn{3}{c}{$r_{\rm res}=2~\rm mm$} 
 & \multicolumn{3}{c}{$r_{\rm res}=5~\rm mm$} \\
\cmidrule(lr){2-4} \cmidrule(lr){5-7} \cmidrule(lr){8-10} \cmidrule(lr){11-13}
 & $P_3$ & $l^{[\Delta]}$ & $L^{[\Delta]}$ 
 & $P_3$ & $l^{[\Delta]}$ & $L^{[\Delta]}$ 
 & $P_3$ & $l^{[\Delta]}$ & $L^{[\Delta]}$ 
 & $P_3$ & $l^{[\Delta]}$ & $L^{[\Delta]}$ \\
\midrule
0    & --      & 0      & 0      & --      & 0      & 0      & --      & 0      & 0      & --      & 0      & 0      \\
0.05   & 0.413   & 52     & 126    & 0.354   & 23     & 65     & 0.447   & 21     & 47     & 0.567   & 17     & 30     \\
0.1    & 0.398   & 208    & 522    & 0.378   & 94     & 249    & 0.360   & 64     & 178    & 0.400   & 46     & 115    \\
0.2    & 0.376   & 742    & 1972   & 0.357   & 350    & 981    & 0.345   & 238    & 689    & 0.384   & 193    & 502    \\
0.4    & 0.371   & 2788   & 7514   & 0.358   & 1361   & 3800   & 0.361   & 932    & 2585   & 0.371   & 696    & 1878   \\
0.6    & 0.349   & 5662   & 16246  & 0.336   & 2810   & 8367   & 0.345   & 1903   & 5520   & 0.357   & 1467   & 4109   \\
0.8    & 0.325   & 9025   & 27729  & 0.310   & 4449   & 14374  & 0.322   & 3008   & 9346   & 0.332   & 2346   & 7056   \\
1.0    & 0.307   & 12684  & 41312  & 0.292   & 6259   & 21418  & 0.305   & 4205   & 13769  & 0.317   & 3270   & 10311  \\
1.4    & 0.276   & 20175  & 73128  & 0.265   & 9989   & 37639  & 0.276   & 6637   & 24034  & 0.286   & 5161   & 18048  \\
1.8    & 0.253   & 27493  & 108457 & 0.248   & 13770  & 55537  & 0.256   & 9031   & 35259  & 0.266   & 7078   & 26594  \\
2.2    & 0.239   & 34312  & 143754 & 0.237   & 17423  & 73396  & 0.246   & 11433  & 46432  & 0.251   & 8774   & 34913  \\
2.6    & 0.229   & 40606  & 177497 & 0.229   & 20754  & 90458  & 0.238   & 13603  & 57128  & 0.240   & 10366  & 43146  \\
3.0    & 0.221   & 45977  & 207687 & 0.224   & 23714  & 105884 & 0.232   & 15570  & 67140  & 0.233   & 11803  & 50555  \\
3.4    & 0.216   & 50671  & 234395 & 0.220   & 26275  & 119514 & 0.228   & 17313  & 76078  & 0.229   & 13140  & 57335  \\
4.0    & 0.211   & 56459  & 267643 & 0.216   & 29581  & 137048 & 0.223   & 19541  & 87717  & 0.224   & 14787  & 66076  \\
5.0    & 0.206   & 63199  & 306134 & 0.212   & 33656  & 158541 & 0.219   & 22463  & 102746 & 0.219   & 17036  & 77613  \\
6.0    & 0.204   & 67154  & 329705 & 0.210   & 36346  & 173415 & 0.215   & 24481  & 113936 & 0.217   & 18802  & 86612  \\
9.0    & 0.200   & 72850  & 363552 & 0.205   & 41253  & 200845 & 0.208   & 28720  & 138321 & 0.213   & 22863  & 107218 \\
12.0   & 0.199   & 75905  & 381240 & 0.204   & 45011  & 220944 & 0.203   & 32300  & 158871 & 0.211   & 26264  & 124341 \\
15.0   & 0.198   & 78361  & 394912 & 0.202   & 48345  & 238988 & 0.201   & 35662  & 177244 & 0.211   & 29463  & 139876 \\
18.0   & 0.198   & 80451  & 406486 & 0.201   & 51422  & 255416 & 0.199   & 38737  & 194469 & 0.210   & 32396  & 154423 \\
\bottomrule
\end{tabular}
\end{table*}

\clearpage

\begin{table*}[!t]
\centering
\caption{Statistics of the alignment degree $P_4$ versus disbalance $\Delta$ using clustering with different cluster sizes $r_{\rm res}$. Centrality class $c=40-75\%$. The alignment degree  $P_4(\Delta)=l^{[\Delta]}/L^{[\Delta]}$, and $L^{[\rm tot]} = 10^6$ denotes the total number of simulated events.}
\label{tab:STAT_4P_40-75}
\begin{tabular}{c|ccc@{\hskip 8pt}|ccc@{\hskip 8pt}|ccc@{\hskip 8pt}|ccc@{\hskip 8pt}}
\toprule
\multirow{2}{*}{$\Delta$,~GeV} 
 & \multicolumn{3}{c}{$r_{\rm res}=0.5~\rm mm$} 
 & \multicolumn{3}{c}{$r_{\rm res}=1~\rm mm$} 
 & \multicolumn{3}{c}{$r_{\rm res}=2~\rm mm$} 
 & \multicolumn{3}{c}{$r_{\rm res}=5~\rm mm$} \\
\cmidrule(lr){2-4} \cmidrule(lr){5-7} \cmidrule(lr){8-10} \cmidrule(lr){11-13}
 & $P_4$ & $l^{[\Delta]}$ & $L^{[\Delta]}$ 
 & $P_4$ & $l^{[\Delta]}$ & $L^{[\Delta]}$ 
 & $P_4$ & $l^{[\Delta]}$ & $L^{[\Delta]}$ 
 & $P_4$ & $l^{[\Delta]}$ & $L^{[\Delta]}$ \\
\midrule
0   & --     & 0    & 0      & --     & 0    & 0      & --     & 0    & 0      & --     & 0    & 0      \\
0.05   & 0.042  & 3    & 72     & 0.027  & 1    & 37     & 0.000  & 0    & 15     & 0.071  & 1    & 14     \\
0.1   & 0.081  & 25   & 309    & 0.063  & 8    & 126    & 0.056  & 4    & 71     & 0.094  & 6    & 64     \\
0.2   & 0.069  & 86   & 1238   & 0.072  & 39   & 541    & 0.067  & 22   & 328    & 0.081  & 22   & 272    \\
0.4   & 0.070  & 352  & 4994   & 0.079  & 189  & 2395   & 0.093  & 132  & 1425   & 0.076  & 85   & 1120   \\
0.6   & 0.073  & 809  & 11023  & 0.083  & 438  & 5286   & 0.088  & 285  & 3227   & 0.073  & 179  & 2467   \\
0.8   & 0.075  & 1429 & 19170  & 0.080  & 741  & 9250   & 0.085  & 475  & 5613   & 0.071  & 308  & 4311   \\
1.0   & 0.073  & 2115 & 29004  & 0.079  & 1100 & 13985  & 0.084  & 710  & 8462   & 0.070  & 459  & 6550   \\
1.4   & 0.068  & 3578 & 52409  & 0.072  & 1816 & 25289  & 0.077  & 1170 & 15196  & 0.065  & 773  & 11816  \\
1.8   & 0.064  & 5101 & 79285  & 0.067  & 2583 & 38381  & 0.072  & 1668 & 23078  & 0.064  & 1135 & 17635  \\
2.2   & 0.061  & 6594 & 107853 & 0.064  & 3317 & 52009  & 0.068  & 2118 & 31168  & 0.061  & 1453 & 23831  \\
2.6   & 0.059  & 8036 & 136137 & 0.062  & 4067 & 65710  & 0.065  & 2561 & 39326  & 0.059  & 1792 & 30118  \\
3.0   & 0.058  & 9375 & 162934 & 0.060  & 4765 & 78794  & 0.063  & 2980 & 47274  & 0.059  & 2128 & 36175  \\
3.4   & 0.056  & 10485& 187775 & 0.059  & 5362 & 91335  & 0.061  & 3363 & 54919  & 0.058  & 2419 & 41973  \\
4.0   & 0.054  & 11929& 220914 & 0.057  & 6149 & 107922 & 0.059  & 3877 & 65360  & 0.056  & 2803 & 49816  \\
5.0   & 0.052  & 13750& 263948 & 0.055  & 7175 & 130413 & 0.057  & 4557 & 80319  & 0.055  & 3387 & 61151  \\
6.0   & 0.051  & 14909& 293625 & 0.054  & 7917 & 147334 & 0.055  & 5061 & 92345  & 0.055  & 3861 & 70027  \\
9.0   & 0.049  & 16643& 337715 & 0.052  & 9366 & 179403 & 0.052  & 6183 & 118788 & 0.054  & 4950 & 91195  \\
12.0  & 0.049  & 17437& 358116 & 0.051  & 10244& 200567 & 0.050  & 7031 & 139818 & 0.054  & 5858 & 108622 \\
15.0  & 0.048  & 18006& 372763 & 0.050  & 11033& 218941 & 0.049  & 7805 & 158445 & 0.054  & 6699 & 124122 \\
18.0  & 0.048  & 18503& 384964 & 0.050  & 11752& 235955 & 0.048  & 8459 & 175819 & 0.054  & 7443 & 138525 \\
\bottomrule
\end{tabular}
\end{table*}

\clearpage

\begin{table*}[!t]
\centering
\caption{Statistics of the alignment degree $P_5$ versus disbalance $\Delta$ using clustering with different cluster sizes $r_{\rm res}$. Centrality class $c=40-75\%$. The alignment degree  $P_5(\Delta)=l^{[\Delta]}/L^{[\Delta]}$, and $L^{[\rm tot]} = 10^6$ denotes the total number of simulated events.}
\label{tab:STAT_5P_40-75}
\begin{tabular}{c|ccc@{\hskip 8pt}|ccc@{\hskip 8pt}|ccc@{\hskip 8pt}|ccc@{\hskip 8pt}}
\toprule
\multirow{2}{*}{$\Delta$,~GeV} 
 & \multicolumn{3}{c}{$r_{\rm res}=0.5~\rm mm$} 
 & \multicolumn{3}{c}{$r_{\rm res}=1~\rm mm$} 
 & \multicolumn{3}{c}{$r_{\rm res}=2~\rm mm$} 
 & \multicolumn{3}{c}{$r_{\rm res}=5~\rm mm$} \\
\cmidrule(lr){2-4} \cmidrule(lr){5-7} \cmidrule(lr){8-10} \cmidrule(lr){11-13}
 & $P_5$ & $l^{[\Delta]}$ & $L^{[\Delta]}$ 
 & $P_5$ & $l^{[\Delta]}$ & $L^{[\Delta]}$ 
 & $P_5$ & $l^{[\Delta]}$ & $L^{[\Delta]}$ 
 & $P_5$ & $l^{[\Delta]}$ & $L^{[\Delta]}$ \\
\midrule
0  & --     & 0     & 0      & --     & 0     & 0      & --     & 0     & 0      & --     & 0     & 0 \\
0.05  & 0.016  & 1     & 62     & 0  & 0     & 31     & 0.111  & 1     & 9      & 0  & 0     & 11 \\
0.1  & 0.017  & 4     & 242    & 0.018  & 2     & 112    & 0.054  & 3     & 56     & 0.024  & 1     & 41 \\
0.2  & 0.014  & 13    & 942    & 0.011  & 5     & 436    & 0.016  & 4     & 246    & 0.021  & 4     & 187 \\
0.4  & 0.019  & 70    & 3615   & 0.019  & 32    & 1666   & 0.019  & 18    & 938    & 0.021  & 15    & 721 \\
0.6  & 0.017  & 142   & 8134   & 0.018  & 67    & 3731   & 0.020  & 40    & 2048   & 0.021  & 34    & 1648 \\
0.8  & 0.017  & 240   & 14107  & 0.019  & 120   & 6451   & 0.022  & 79    & 3592   & 0.021  & 59    & 2832 \\
1.0  & 0.017  & 372   & 21526  & 0.018  & 179   & 9841   & 0.022  & 119   & 5480   & 0.020  & 85    & 4256 \\
1.4  & 0.017  & 665   & 39647  & 0.019  & 335   & 17961  & 0.019  & 193   & 9904   & 0.018  & 145   & 7848 \\
1.8  & 0.016  & 992   & 61074  & 0.018  & 490   & 27574  & 0.018  & 278   & 15181  & 0.018  & 214   & 12055 \\
2.2  & 0.015  & 1295  & 84399  & 0.017  & 649   & 37918  & 0.017  & 357   & 21010  & 0.016  & 267   & 16645 \\
2.6  & 0.015  & 1625  & 108193 & 0.016  & 804   & 48914  & 0.017  & 457   & 27205  & 0.016  & 332   & 21379 \\
3.0  & 0.014  & 1869  & 131650 & 0.016  & 938   & 59747  & 0.016  & 548   & 33435  & 0.015  & 402   & 26300 \\
3.4  & 0.014  & 2126  & 154614 & 0.015  & 1074  & 70207  & 0.016  & 636   & 39529  & 0.015  & 457   & 30913 \\
4.0  & 0.013  & 2495  & 186146 & 0.015  & 1264  & 84983  & 0.015  & 741   & 48250  & 0.014  & 533   & 37621 \\
5.0  & 0.013  & 2931  & 228822 & 0.014  & 1513  & 106106 & 0.015  & 916   & 61283  & 0.014  & 665   & 47413 \\
6.0  & 0.013  & 3269  & 260665 & 0.014  & 1698  & 122922 & 0.014  & 1047  & 72540  & 0.014  & 783   & 55860 \\
9.0  & 0.012  & 3768  & 312789 & 0.013  & 2038  & 156987 & 0.013  & 1290  & 98952  & 0.013  & 1010  & 75636 \\
12.0 & 0.012  & 3973  & 336032 & 0.013  & 2274  & 179439 & 0.012  & 1477  & 119982 & 0.013  & 1238  & 92108 \\
15.0 & 0.012  & 4091  & 351362 & 0.012  & 2427  & 198116 & 0.012  & 1635  & 138576 & 0.013  & 1450  & 107582 \\
18.0 & 0.011  & 4182  & 363668 & 0.012  & 2587  & 215213 & 0.012  & 1796  & 155784 & 0.014  & 1649  & 121979 \\
\bottomrule
\end{tabular}
\end{table*}

\newpage

\subsection{Centrality 0\%-75\%}
\label{Stats0-75}

\begin{table*}[b]
\centering
\caption{Statistics of the alignment degree $P_3$ versus disbalance $\Delta$ using clustering with different cluster sizes $r_{\rm res}$. Centrality class $c=0-75\%$. The alignment degree  $P_3(\Delta)=l^{[\Delta]}/L^{[\Delta]}$, and $L^{[\rm tot]} = 10^6$ denotes the total number of simulated events.}
\label{tab:STAT_3P_0-75}
\begin{tabular}{c|ccc@{\hskip 8pt}|ccc@{\hskip 8pt}|ccc@{\hskip 8pt}|ccc@{\hskip 8pt}}
\toprule
\multirow{2}{*}{$\Delta$, GeV} 
 & \multicolumn{3}{c}{$r_{\rm res}=0.5~\rm mm$} 
 & \multicolumn{3}{c}{$r_{\rm res}=1~\rm mm$} 
 & \multicolumn{3}{c}{$r_{\rm res}=2~\rm mm$} 
 & \multicolumn{3}{c}{$r_{\rm res}=5~\rm mm$} \\
\cmidrule(lr){2-4} \cmidrule(lr){5-7} \cmidrule(lr){8-10} \cmidrule(lr){11-13}
 & $P_3$ & $l^{[\Delta]}$ & $L^{[\Delta]}$ 
 & $P_3$ & $l^{[\Delta]}$ & $L^{[\Delta]}$ 
 & $P_3$ & $l^{[\Delta]}$ & $L^{[\Delta]}$ 
 & $P_3$ & $l^{[\Delta]}$ & $L^{[\Delta]}$ \\
\midrule
0  & --     & 0      & 0      & --     & 0      & 0      & --     & 0      & 0      & --     & 0      & 0 \\
0.05  & 0.338  & 48     & 142    & 0.364  & 24     & 66     & 0.242  & 8      & 33     & 0.286  & 8      & 28 \\
0.1  & 0.397  & 215    & 541    & 0.399  & 119    & 298    & 0.431  & 78     & 181    & 0.377  & 58     & 154 \\
0.2  & 0.384  & 751    & 1958   & 0.384  & 404    & 1051   & 0.406  & 274    & 675    & 0.420  & 237    & 564 \\
0.4  & 0.370  & 2745   & 7418   & 0.360  & 1410   & 3920   & 0.372  & 961    & 2580   & 0.388  & 785    & 2023 \\
0.6  & 0.350  & 5637   & 16126  & 0.332  & 2789   & 8393   & 0.343  & 1862   & 5432   & 0.364  & 1542   & 4240 \\
0.8  & 0.331  & 9097   & 27520  & 0.315  & 4534   & 14377  & 0.324  & 2993   & 9229   & 0.342  & 2457   & 7187 \\
1.0  & 0.311  & 12900  & 41413  & 0.298  & 6412   & 21485  & 0.310  & 4230   & 13665  & 0.316  & 3339   & 10578 \\
1.4  & 0.279  & 20431  & 73155  & 0.271  & 10272  & 37846  & 0.281  & 6711   & 23877  & 0.286  & 5218   & 18245 \\
1.8  & 0.255  & 27579  & 107983 & 0.251  & 13956  & 55563  & 0.259  & 9059   & 34915  & 0.263  & 6983   & 26575 \\
2.2  & 0.240  & 34353  & 143110 & 0.239  & 17533  & 73377  & 0.246  & 11302  & 45937  & 0.249  & 8713   & 34966 \\
2.6  & 0.230  & 40563  & 176589 & 0.231  & 20820  & 90325  & 0.235  & 13328  & 56651  & 0.239  & 10270  & 42887 \\
3.0  & 0.222  & 45980  & 206933 & 0.225  & 23781  & 105723 & 0.230  & 15278  & 66402  & 0.232  & 11699  & 50368 \\
3.4  & 0.218  & 50760  & 233367 & 0.221  & 26420  & 119554 & 0.225  & 16987  & 75348  & 0.228  & 13009  & 57135 \\
4.0  & 0.212  & 56520  & 266037 & 0.217  & 29633  & 136767 & 0.221  & 19120  & 86707  & 0.223  & 14742  & 65981 \\
5.0  & 0.207  & 62975  & 304101 & 0.212  & 33574  & 158092 & 0.215  & 21881  & 101667 & 0.218  & 16938  & 77586 \\
6.0  & 0.205  & 67050  & 327704 & 0.210  & 36293  & 172806 & 0.212  & 24005  & 112986 & 0.216  & 18681  & 86611 \\
9.0  & 0.201  & 72672  & 361356 & 0.207  & 41386  & 200390 & 0.206  & 28409  & 137800 & 0.212  & 22728  & 107161 \\
12.0 & 0.200  & 75800  & 379143 & 0.204  & 45063  & 220707 & 0.203  & 32003  & 157919 & 0.210  & 26144  & 124636 \\
15.0 & 0.199  & 78204  & 392852 & 0.203  & 48320  & 238545 & 0.200  & 35276  & 176282 & 0.209  & 29279  & 140393 \\
18.0 & 0.199  & 80355  & 404720 & 0.202  & 51443  & 254959 & 0.199  & 38332  & 192999 & 0.208  & 32225  & 154945 \\
\bottomrule
\end{tabular}
\end{table*}
\clearpage

\begin{table*}[!b]
\centering
\caption{Statistics of the alignment degree $P_4$ versus disbalance $\Delta$ using clustering with different cluster sizes $r_{\rm res}$. Centrality class $c=0-75\%$. The alignment degree  $P_4(\Delta)=l^{[\Delta]}/L^{[\Delta]}$, and $L^{[\rm tot]} = 10^6$ denotes the total number of simulated events.}
\label{tab:STAT_4P_0-75}
\begin{tabular}{c|ccc@{\hskip 8pt}|ccc@{\hskip 8pt}|ccc@{\hskip 8pt}|ccc@{\hskip 8pt}}
\toprule
\multirow{2}{*}{$\Delta$, GeV} 
 & \multicolumn{3}{c}{$r_{\rm res}=0.5~\rm mm$} 
 & \multicolumn{3}{c}{$r_{\rm res}=1~\rm mm$} 
 & \multicolumn{3}{c}{$r_{\rm res}=2~\rm mm$} 
 & \multicolumn{3}{c}{$r_{\rm res}=5~\rm mm$} \\
\cmidrule(lr){2-4} \cmidrule(lr){5-7} \cmidrule(lr){8-10} \cmidrule(lr){11-13}
 & $P_4$ & $l^{[\Delta]}$ & $L^{[\Delta]}$ 
 & $P_4$ & $l^{[\Delta]}$ & $L^{[\Delta]}$ 
 & $P_4$ & $l^{[\Delta]}$ & $L^{[\Delta]}$ 
 & $P_4$ & $l^{[\Delta]}$ & $L^{[\Delta]}$ \\
\midrule
0  & --     & 0      & 0      & --     & 0      & 0      & --     & 0      & 0      & --     & 0      & 0 \\
0.05  & 0.064  & 3      & 47     & 0.059  & 1      & 17     & 0.118  & 2      & 17     & 0  & 0      & 15 \\
0.1  & 0.087  & 18     & 208    & 0.113  & 11     & 97     & 0.133  & 11     & 83     & 0.068  & 5      & 74 \\
0.2  & 0.071  & 55     & 777    & 0.076  & 30     & 395    & 0.063  & 20     & 316    & 0.098  & 25     & 256 \\
0.4  & 0.077  & 233    & 3027   & 0.084  & 132    & 1566   & 0.079  & 88     & 1113   & 0.089  & 86     & 962 \\
0.6  & 0.074  & 485    & 6593   & 0.077  & 265    & 3424   & 0.082  & 194    & 2356   & 0.078  & 156    & 1997 \\
0.8  & 0.072  & 822    & 11406  & 0.073  & 425    & 5826   & 0.080  & 315    & 3937   & 0.077  & 262    & 3384 \\
1.0  & 0.070  & 1209   & 17372  & 0.071  & 621    & 8722   & 0.077  & 452    & 5899   & 0.073  & 368    & 5057 \\
1.4  & 0.067  & 2113   & 31563  & 0.069  & 1090   & 15880  & 0.071  & 754    & 10563  & 0.067  & 605    & 9006 \\
1.8  & 0.063  & 3004   & 47437  & 0.066  & 1567   & 23680  & 0.068  & 1066   & 15755  & 0.066  & 875    & 13323 \\
2.2  & 0.061  & 3898   & 64361  & 0.063  & 2026   & 32066  & 0.065  & 1369   & 21116  & 0.063  & 1118   & 17875 \\
2.6  & 0.058  & 4704   & 81030  & 0.061  & 2453   & 40252  & 0.063  & 1658   & 26453  & 0.061  & 1358   & 22347 \\
3.0  & 0.056  & 5430   & 96816  & 0.060  & 2879   & 48186  & 0.062  & 1957   & 31804  & 0.059  & 1592   & 26779 \\
3.4  & 0.055  & 6091   & 111625 & 0.058  & 3261   & 55865  & 0.060  & 2227   & 36912  & 0.058  & 1785   & 30911 \\
4.0  & 0.053  & 6951   & 130924 & 0.057  & 3748   & 65838  & 0.059  & 2564   & 43783  & 0.056  & 2052   & 36579 \\
5.0  & 0.051  & 7944   & 155853 & 0.054  & 4316   & 79277  & 0.057  & 3034   & 53631  & 0.054  & 2407   & 44329 \\
6.0  & 0.050  & 8631   & 173045 & 0.053  & 4787   & 89503  & 0.055  & 3387   & 61278  & 0.054  & 2722   & 50618 \\
9.0  & 0.048  & 9610   & 199445 & 0.051  & 5598   & 109072 & 0.052  & 4088   & 78198  & 0.053  & 3424   & 64477 \\
12.0 & 0.047  & 10092  & 212627 & 0.050  & 6143   & 122667 & 0.050  & 4585   & 91434  & 0.053  & 3973   & 75361 \\
15.0 & 0.047  & 10411  & 222183 & 0.049  & 6595   & 134554 & 0.049  & 5039   & 103116 & 0.052  & 4439   & 85212 \\
18.0 & 0.047  & 10740  & 230606 & 0.048  & 7011   & 145341 & 0.048  & 5500   & 114024 & 0.052  & 4909   & 94301 \\
\bottomrule
\end{tabular}
\end{table*}
\clearpage

\begin{table*}[!t]
\centering
\caption{Statistics of the alignment degree $P_5$ versus disbalance $\Delta$ using clustering with different cluster sizes $r_{\rm res}$. Centrality class $c=0-75\%$. The alignment degree  $P_5(\Delta)=l^{[\Delta]}/L^{[\Delta]}$, and $L^{[\rm tot]} = 10^6$ denotes the total number of simulated events.}
\label{tab:STAT_5P_0-75}
\begin{tabular}{c|ccc@{\hskip 8pt}|ccc@{\hskip 8pt}|ccc@{\hskip 8pt}|ccc@{\hskip 8pt}}
\toprule
\multirow{2}{*}{$\Delta$, GeV} 
 & \multicolumn{3}{c}{$r_{\rm res}=0.5~\rm mm$} 
 & \multicolumn{3}{c}{$r_{\rm res}=1~\rm mm$} 
 & \multicolumn{3}{c}{$r_{\rm res}=2~\rm mm$} 
 & \multicolumn{3}{c}{$r_{\rm res}=5~\rm mm$} \\
\cmidrule(lr){2-4} \cmidrule(lr){5-7} \cmidrule(lr){8-10} \cmidrule(lr){11-13}
 & $P_5$ & $l^{[\Delta]}$ & $L^{[\Delta]}$ 
 & $P_5$ & $l^{[\Delta]}$ & $L^{[\Delta]}$ 
 & $P_5$ & $l^{[\Delta]}$ & $L^{[\Delta]}$ 
 & $P_5$ & $l^{[\Delta]}$ & $L^{[\Delta]}$ \\
\midrule
0  & --     & 0      & 0      & --     & 0      & 0      & --     & 0      & 0      & --     & 0      & 0 \\
0.05  & 0  & 0      & 33     & 0  & 0      & 16     & 0  & 0      & 7      & 0  & 0      & 13 \\
0.1  & 0.014  & 2      & 142    & 0  & 0      & 68     & 0.024  & 1      & 41     & 0.024  & 1      & 42 \\
0.2  & 0.022  & 13     & 581    & 0.014  & 4      & 277    & 0.030  & 5      & 166    & 0.026  & 4      & 156 \\
0.4  & 0.022  & 49     & 2267   & 0.019  & 21     & 1089   & 0.033  & 20     & 609    & 0.019  & 12     & 621 \\
0.6  & 0.018  & 86     & 4834   & 0.017  & 39     & 2289   & 0.027  & 38     & 1415   & 0.018  & 25     & 1363 \\
0.8  & 0.019  & 159    & 8590   & 0.017  & 69     & 4024   & 0.023  & 57     & 2475   & 0.020  & 48     & 2365 \\
1.0  & 0.018  & 231    & 13042  & 0.016  & 99     & 6143   & 0.020  & 75     & 3782   & 0.019  & 67     & 3557 \\
1.4  & 0.017  & 398    & 24104  & 0.017  & 187    & 11330  & 0.018  & 130    & 7070   & 0.017  & 108    & 6337 \\
1.8  & 0.016  & 578    & 36961  & 0.016  & 271    & 17218  & 0.016  & 179    & 10850  & 0.015  & 151    & 9743 \\
2.2  & 0.015  & 748    & 50688  & 0.016  & 370    & 23724  & 0.015  & 225    & 14960  & 0.016  & 206    & 13184 \\
2.6  & 0.014  & 920    & 64853  & 0.015  & 452    & 30210  & 0.015  & 280    & 19096  & 0.015  & 252    & 16780 \\
3.0  & 0.014  & 1067   & 78751  & 0.014  & 525    & 36910  & 0.014  & 323    & 23447  & 0.014  & 287    & 20364 \\
3.4  & 0.013  & 1225   & 92033  & 0.014  & 617    & 43303  & 0.013  & 371    & 27587  & 0.014  & 326    & 23809 \\
4.0  & 0.013  & 1425   & 110252 & 0.014  & 721    & 52200  & 0.013  & 435    & 33321  & 0.013  & 368    & 28705 \\
5.0  & 0.012  & 1682   & 135371 & 0.013  & 874    & 64871  & 0.013  & 552    & 41950  & 0.013  & 458    & 35924 \\
6.0  & 0.012  & 1863   & 153818 & 0.013  & 982    & 75050  & 0.013  & 635    & 49516  & 0.013  & 539    & 41782 \\
9.0  & 0.012  & 2152   & 184339 & 0.013  & 1214   & 95679  & 0.012  & 811    & 66346  & 0.013  & 709    & 55202 \\
12.0 & 0.011  & 2266   & 198635 & 0.012  & 1342   & 109683 & 0.012  & 952    & 79797  & 0.013  & 847    & 65910 \\
15.0 & 0.011  & 2370   & 208770 & 0.012  & 1449   & 121802 & 0.012  & 1055   & 91605  & 0.012  & 942    & 75743 \\
18.0 & 0.011  & 2443   & 217326 & 0.012  & 1560   & 132835 & 0.011  & 1144   & 102445 & 0.012  & 1049   & 84793 \\
\bottomrule
\end{tabular}
\end{table*}
\clearpage

\bibliographystyle{apsrev4-2}
\bibliography{REF_ALIGN_25_26}

\end{document}